\documentclass[twocolumn,pra,aps,showpacs,superscriptaddress]{revtex4-1}
\usepackage[utf8]{inputenc}
\usepackage[T1]{fontenc}
\usepackage{hyperref}
\usepackage{graphicx}
\usepackage{bbm}
\usepackage{bm}
\usepackage[caption=false]{subfig}
\usepackage{color}
\usepackage{dsfont}
\usepackage{bigints}

\DeclareMathOperator{\tr}{tr}

\newcommand{\phid}{\phi^\dag}
\newcommand{\psid}{\psi^\dag}

\newcommand{\VPHI}{\bm{\varphi}}
\newcommand{\PHI}{\bm{\phi}}

\newcommand{\MR}{\mathbf{R}}
\newcommand{\MG}{\mathbf{G}}
\newcommand{\MGamma}{\bm{\Gamma}}
\newcommand{\Q}{\mathbf{q}}
\newcommand{\X}{\mathbf{x}}
\newcommand{\vP}{\mathbf{p}}

\usepackage{hyperref}
\hypersetup{
   pdfpagemode=UseNone, 
   pdfstartpage=1,
   pdfmenubar=true,
   pdftoolbar=true,
   colorlinks = true,
   linkcolor=blue,
   citecolor=blue,
   urlcolor=blue,
   bookmarksopen=false
 }

\begin{document}

\title{Renormalization group study of Bose polarons}
\author{Felipe Isaule}
\affiliation{Departament de F\'isica Qu\`antica i Astrof\'isica, 
Facultat de F\'{\i}sica, and Institut de Ci\`encies del Cosmos (ICCUB), Universitat de Barcelona, 
Mart\'i i Franqu\`es 1, E–08028 Barcelona, Spain}
\author{Ivan Morera}
\affiliation{Departament de F\'isica Qu\`antica i Astrof\'isica, 
Facultat de F\'{\i}sica, and Institut de Ci\`encies del Cosmos (ICCUB), Universitat de Barcelona, 
Mart\'i i Franqu\`es 1, E–08028 Barcelona, Spain}
\author{Pietro Massignan}
\affiliation{Departament de F\'isica, Universitat Polit\`ecnica de Catalunya, Campus Nord B4-B5, E-08034 Barcelona, Spain}
\author{Bruno Juli\'{a}-D\'{i}az}
\affiliation{Departament de F\'isica Qu\`antica i Astrof\'isica, 
Facultat de F\'{\i}sica, and Institut de Ci\`encies del Cosmos (ICCUB), Universitat de Barcelona, 
Mart\'i i Franqu\`es 1, E–08028 Barcelona, Spain}

\date{\today}
\begin{abstract}
We study the properties of a single impurity in a dilute Bose gas, a Bose polaron, using the functional renormalization group. We use an ansatz for the effective action motivated by a derivative expansion, and we compute the energies of the attractive and repulsive branches of excitations in both two and three spatial dimensions. Three-body correlations play an important role in the attractive branch, and we account for those by including three-body couplings between two bath bosons and the impurity. Our calculations compare very favorably with state-of-the-art experimental measurements and numerical simulations.
\end{abstract}
\maketitle

\section{Introduction}

The study of an impurity immersed in a quantum medium has a long history, dating back to the work of Landau and Pekar on electrons coupled to an ionic crystal~\cite{landau_effective_1948}. Such impurity is understood as a  dressed quasiparticle referred to as a polaron.
Impurities in bosonic baths have been studied in a variety of configurations and played a key role in elucidating the physics of helium liquids
\cite{Fabrocini1986,Fabrocini1998}.
The polaron problem has been extensively studied also in fermionic mediums, particularly in condensed matter  physics~\cite{alexandrov_advances_2010} and ultracold atoms~\cite{Schirotzek2009oof,Nascimbene2009coo,Kohstall2012mac,Massignan2014,Levinsen2015,Cetina2015doi,Cetina2016umb,Scazza2017rfp,DarkwahOppong2019ooc,Ness2020ooa,Adlong2020,Fritsche2021}. 

The interest in an impurity immersed in a weakly interacting Bose gas, the Bose polaron, has greatly increased in the past decade. Indeed, the experimental progress in cold atom gases~\cite{bloch_many-body_2008} recently allowed the experimental realization of Bose polarons, including the regime of resonant boson-impurity  interactions~\cite{catani_quantum_2012,jorgensen_observation_2016,hu_bose_2016,yan_bose_2020}.
Theoretically, early perturbative works were restricted to the regime of weak boson-impurity interactions~\cite{astrakharchik_motion_2004,tempere_feynman_2009,levinsen_finite-temperature_2017}. Improved descriptions emerged in recent years with a variety of techniques, including field theory approaches~\cite{rath_field-theoretical_2013,camacho-guardian_landau_2018,pastukhov_polaron_2018,pastukhov_polaron_2018-1,guenther_bose_2018}, variational methods~\cite{Massignan2005,li_variational_2014,levinsen_impurity_2015,shchadilova_quantum_2016,yoshida_universality_2018,dehkharghani_coalescence_2018,drescher_real-space_2019,guenther_mobile_2021,massignan_universal_2021,pena_ardila_dynamical_2021,levinsen_quantum_2021,schmidt_self-stabilized_2021}, the Fr\"ohlich model~\cite{grusdt_renormalization_2015,grusdt_strong-coupling_2017,ichmoukhamedov_feynman_2019}, and Monte Carlo (MC) simulations~\cite{pena_ardila_impurity_2015,pena_ardila_bose_2016,pena_ardila_analyzing_2019,pena_ardila_strong_2020,pascual_quasiparticle_2021}.

In the case of Bose polarons, the bosonic nature of the medium means that three- and more-body interactions can be important~\cite{levinsen_impurity_2015}, and that the mixture is generally highly unstable against three-body losses. In contrast, in the case of Fermi polarons, interactions beyond the two-body level are suppressed by Pauli blocking, and the mixture is relatively long-lived. These differences make the theoretical description and experimental investigation of Bose polarons much more challenging. Furthermore, in a bosonic bath the dressed impurity 
has the same quantum statistics in the two asymptotic limits of zero and 
infinite attraction, so that the problem features a smooth polaron-to-molecule crossover~\cite{rath_field-theoretical_2013}. In contrast, in a fermionic medium, a bare impurity and the molecule it forms by binding to a bath fermion have opposite quantum statistics, and therefore the spectrum of a Fermi polaron generally features a sharp polaron-to-molecule transition. 

In this work we study  Bose polarons using the functional renormalization group (FRG) 
based on the effective average action~\cite{wetterich_exact_1993,berges_non-perturbative_2002} 
(for a complete review see Ref.~\cite{dupuis_nonperturbative_2021}). The FRG is a non-perturbative field theory approach which has proved to be a powerful tool to study strongly correlated systems, including Efimov physics in three-body~\cite{moroz_efimov_2009,floerchinger_functional_2009,floerchinger_efimov_2011} and four-body~\cite{schmidt_renormalization-group_2010,jaramillo_avila_universal_2013,jaramillo_avila_four-boson_2015,horinouchi_topological_2016} systems, the BCS-BEC crossover~\cite{birse_pairing_2005,diehl_universality_2006,diehl_functional_2007,floerchinger_particle-hole_2008,floerchinger_modified_2010,diehl_functional_2010,boettcher_critical_2014,boettcher_phase_2015,faigle-cedzich_dimensional_2021}, and the Fermi polaron~\cite{schmidt_excitation_2011,kamikado_mobile_2017,von_milczewski_functional_2021} (for applications in other areas of physics see Refs.~\cite{berges_non-perturbative_2002,dupuis_nonperturbative_2021}). 
The FRG permits us to include systematically the effect of fluctuations, such as those arising from three- and more-body correlations,  and to add their effect non-perturbatively over a wide range of scales. In addition, the FRG already provided accurate descriptions of Bose gases in two and three dimensions~\cite{dupuis_non-perturbative_2007,wetterich_functional_2008,floerchinger_functional_2008,floerchinger_superfluid_2009,dupuis_infrared_2009,isaule_thermodynamics_2020,isaule_functional_2021}, including critical phenomena at the superfluid phase transitions~\cite{blaizot_non-perturbative_2005}. Therefore, the FRG appears to be good technique to study novel physics in Bose polarons. 
Furthermore, because the FRG provides a unified description of  few- and many-body physics within the same theory, studying polaron physics with FRG can provide important insight into Bose-Bose and Bose-Fermi mixtures.

In this article, we focus on Bose polarons at zero temperature in two and three spatial dimensions. We approximate the coarse-grained effective action under a derivative expansion, and we consider up to three-body correlations. Our approximation enables us to give a good description of the ground state energies and to quantify the importance of three-body forces.
This article is organized as follows. 
In Sec.~\ref{sec:model} we present our model and introduce the FRG flow equation. 
In Sec.~\ref{sec:RP} we study the repulsive branch of the Bose polaron, presenting the main aspects of our FRG calculations, as well as results for the polaron energy. 
In Sec.~\ref{sec:AP} we study the attractive branch, stressing the specific considerations for the study of attractive interactions, and presenting results for the polaron energy with and without three-body correlations. 
In Sec.~\ref{sec:concl} we present the conclusions and outlook of our work. 
Finally, in Appendixes~\ref{app:RP} and \ref{app:AP} we provide specific details of the RG equations, and in Appendix~\ref{app:meff} we provide an estimation for the effective mass in the repulsive branch.

\section{Model and FRG equation}
\label{sec:model}

We consider an impurity of mass $m_I$ and energy $\mu_I$ immersed in a gas of weakly repulsive bosons of mass $m_B$ and chemical potential $\mu_B$. We 
approximate the boson-boson and boson-impurity interactions with contact potentials of strengths $g_{BB}$ and $g_{BI}$, respectively. In a field theory formulation, such a system is described by the microscopic
action~\cite{rath_field-theoretical_2013}
\begin{align}
    \mathcal{S}[\VPHI]=\int_x &\bigg[\psid_B\left(\partial_\tau-\frac{\nabla^2}{2m_B}-\mu_B\right)\psi_B\nonumber\\
    &+\psid_I\left(\partial_\tau-\frac{\nabla^2}{2m_I}-\mu_I\right)\psi_I\nonumber\\
    &+\frac{g_{BB}}{2}(\psid_B\psi_B)^2+g_{BI}\psid_B\psid_I\psi_B\psi_I\bigg]\,,
    \label{sec:model;eq:S}
\end{align}
where we use natural units $\hbar=1$, and $\int_x=\int_0^\infty d\tau\int d\X$, with 
$\tau=i t$ being the imaginary time. The microscopic action defines the grand-canonical partition function~\cite{berges_non-perturbative_2002}, and is a functional of the complex fields 
$\psi_B$ and $\psi_I$, which represent the bath bosons and the impurity, respectively. Since we consider a single impurity, the quantum statistics of $\psi_I$ is irrelevant. 

The boson-boson interaction needs to be repulsive, in order to prevent the collapse of the bosonic medium. In contrast, the boson-impurity interaction can either be repulsive or attractive, 
leading to the repulsive and attractive branches of the Bose polaron. 
The repulsive branch is generally well described by perturbative approaches. 
In contrast, the attractive branch is more challenging to describe and shows richer physics. In particular, for strong attractive coupling the scattering length diverges in three dimensions, so that usual perturbative approaches may not be employed.  
Furthermore, three- and more-body physics can become important in the regime of strong coupling~\cite{levinsen_impurity_2015}. Therefore, more robust approaches have to be employed in this regime~\cite{yoshida_universality_2018}.
An analytic solution for the case of heavy polarons at unitarity was recently put forward in Ref.~\cite{massignan_universal_2021}.

In this work, we extract the ground state properties of the Bose polaron from the 
Green's functions~\cite{rath_field-theoretical_2013}.
These can be obtained from $\mathcal{S}$ by taking into account all the quantum paths using the \emph{path integral} formalism. However, it is more convenient to work in terms of the Legendre-transformed effective action $\Gamma$. The effective action is defined in terms of classical fields, and thus it already contains the effect of fluctuations. The Green's functions are then naturally obtained from the vertex functions $\Gamma^{(n)}$ (for details see Ref.~\cite{berges_non-perturbative_2002}).

The effective action can be calculated perturbatively from a loop expansion. However, this is impractical in the regime of strong coupling, where one needs to take into account fluctuations over a wide range of scales. 
Within the FRG, the calculation of $\Gamma$ is instead performed non-perturbatively. In this framework, a regulator function $\MR_k$ is added to the theory to suppress fluctuations at momenta $q\lesssim k$, so one works in terms of a $k$-dependent effective action $\Gamma_k$. 
At a high scale in the ultraviolet (UV)
$k=\Lambda$, \emph{all} fluctuations are suppressed, and the effective action is simply 
the microscopic action $\Gamma_\Lambda=\mathcal{S}$. On the other hand, for 
$k\to 0$ all fluctuations are considered, and $\Gamma_0$ is the full effective action.

The flow of $\Gamma_k$ as a function of $k$ is dictated by the Wetterich equation~\cite{wetterich_exact_1993}
\begin{equation}
\partial_k \Gamma_k=\frac{1}{2}\tr\left[(\MGamma_k^{(2)}+\MR_k)^{-1}\partial_k \MR_k\right]\,,
\label{sec:model;eq:WettEq}
\end{equation}
where $\MGamma^{(2)}_k$ is the matrix with the second functional derivatives of $\Gamma_k$,
\begin{equation}
    \MGamma_k^{(2)}=\frac{\delta^2\Gamma_k}{\delta\VPHI_{-q}^\dag\delta\VPHI_q}\,,
\end{equation}
and tr denotes both a matrix trace and an 
integral over internal momentum $q=(\omega,\Q)$. The Wetterich equation has a one loop structure with a propagator $\MG_k=(\MGamma_k^{(2)}+\MR_k)^{-1}$, and insertion $\partial_k\MR_k$~\cite{wetterich_exact_1993}.

In most applications, one solves the RG flow by proposing an ansatz for $\Gamma_k$, which respects the symmetries of the microscopic theory. In this 
work we employ an ansatz based on a derivative expansion (DE) truncated to a small number of $k$-dependent couplings~\cite{dupuis_nonperturbative_2021}. 
Within the DE, we expand the effective action up to a chosen number of fields and derivatives, so the Wetterich equation becomes a set of coupled differential equations for the $k$-dependent couplings in the expansion. These equations can then be solved numerically using standard methods.

It has been shown that the DE gives an accurate description of various properties of the Fermi polaron in both two~\cite{von_milczewski_functional_2021} and three~\cite{schmidt_excitation_2011} dimensions, including the onset of the polaron and molecule phases and their respective energies. Similarly, in this work we show that the DE provides a precise description of the ground state energy of the Bose polaron and also enables us to quantify the importance of three-body correlations.

In the following, we present the study of the repulsive and attractive branches separately. We start in Sec.~\ref{sec:RP} with the simpler repulsive branch to easily introduce our formalism. We then generalize our formalism to the attractive branch in Sec.~\ref{sec:AP}. 

\section{Repulsive Bose polarons}
\label{sec:RP}

We start from action~(\ref{sec:model;eq:S}), and we neglect
the feedback of the impurity on the medium. To solve the RG flow of the effective action in the presence of repulsive impurity-bath interactions, we propose the following ansatz
\begin{align}
    \Gamma_k[\VPHI]=&\int_x \bigg[\psid_B\left(S_B\partial_\tau-\frac{Z_B}{2m_B}\nabla^2-V_B\partial^2_\tau\right)\psi_B\nonumber\\
    &+\psid_I\left(S_I\partial_\tau-\frac{Z_I}{2m_I}\nabla^2\right)\psi_I+U(\rho_B,\rho_I)\bigg]\,,
    \label{sec:RP;eq:Gamma}
\end{align}
where $S_B$, $Z_B$, $V_B$, $S_I$, and $Z_I$ are renormalization factors which we assume are field-independent, and
\begin{equation}
    U=-P+u_I \rho_I+\frac{\lambda_{BB}}{2}(\rho_B-\rho_0)^2+\lambda_{BI}(\rho_B-\rho_0)\rho_I\,,
    \label{sec:RP;eq:U}
\end{equation}
is the effective potential expanded up to fourth-order in the fields, where $\rho_a=\psid_a\psi_a$ ($a=B,I$), $P$ is the pressure of the bath, $u_I$ is a one-body coupling for the impurity, and $\lambda_{BB}$ and 
$\lambda_{BI}$ are the couplings associated with the boson-boson and boson-impurity 
interactions, respectively. All the couplings in the expansion, $S_B$, $Z_B$, $V_B$, $S_I$, $Z_I$, $P$, $u_I$, $\lambda_{BB}$, and $\lambda_{BI}$, as well as the order parameter $\rho_0=\langle\rho_B\rangle$, flow with $k$. We note that three- and more-body correlations are not important in the repulsive branch, and thus ansatz~(\ref{sec:RP;eq:Gamma}) only contains two-body couplings.

Our ansatz is based on the one used to study repulsive Bose-Bose mixtures~\cite{isaule_functional_2021}, adapted to the limit of extreme population imbalance. The term  $V_B \partial_\tau^2$ is necessary to correctly describe the bosonic medium in the infrared where the effective action develops phonons with linear dispersion, taking the form of a relativistic model~\cite{wetterich_functional_2008}. An analogous term is not needed for the impurity, as the latter is not condensed. 

We stress that our ansatz is accurate for only two- and three-dimensional gases. In one dimension, our level of truncation is not able to accurately capture the quasicondensate nature of the bath~\cite{dupuis_non-perturbative_2007}, where we have to carefully treat the stronger impact of phase fluctuations~\cite{isaule_thermodynamics_2020}.

The minimum $\rho_0$ of the effective potential corresponds to the condensate density of the medium, giving its physical value at $k=0$. If $\rho_0>0$, the $U(1)$-symmetry of the bosonic bath is broken, and the gas is condensed. Here we study the two- and three-dimensional polarons at zero temperature, so that $\rho_0$ is always non-zero. In contrast, for the impurity $\langle\rho_I\rangle=0$. Furthermore, the superfluid density 
is given by the value at $k=0$ of the superfluid stiffness $\rho_s=Z_B\rho_0$~\cite{dupuis_infrared_2009}. Because at zero temperature all bosons 
are superfluid, we can extract the density of the medium $n$ from $n=\rho_s$~\cite{floerchinger_functional_2008}. 
We note that interactions deplete the condensate, $\rho_0\leq\rho_s=n$, and therefore, the mass renormalization coefficient $Z_B$ flows to a value larger than unity for $k\to 0$~\cite{floerchinger_functional_2008}.

To solve the RG flow we need an equation for each running coupling. We obtain the flow equations from the Wetterich equation~(\ref{sec:model;eq:WettEq}). They can be found in App.~\ref{app:RP}. In addition, we need to choose a regulator. In this work, we use the optimized Litim 
regulator~\cite{litim_optimisation_2000}
\begin{equation}
    R_{k,a}=\frac{Z_a}{2m_a}(k^2-\Q^2)\Theta(k^2-\Q^2)\,,\qquad a=B,I
    \label{sec:RP;eq:Rlitim}
\end{equation}
where $\Theta$ is the Heaviside step function. This choice enables us to perform 
the momentum integrals analytically before solving the RG flow.
Finally, we need to specify the initial conditions of the RG flow. We do so in the following subsection.

\subsection{Initial conditions of the RG flow}
\label{sec:RP;sub:IC}

The RG flow is started at a scale $k=\Lambda$ much larger than the relevant scale of the bath, which, in this case, is given by the healing scale $k_{h}=(2m_B\mu_B)^{1/2}$~\cite{floerchinger_functional_2008,isaule_thermodynamics_2020}. At this high scale, we 
can impose that $\Gamma_\Lambda=\mathcal{S}$. We obtain
\begin{align}
    &S_B(\Lambda)=Z_B(\Lambda)=S_I(\Lambda)=Z_I(\Lambda)=1\,,\quad V_B(\Lambda)=0\,,\nonumber\\ 
    &\rho_0(\Lambda)=\frac{\mu_B}{\lambda_{BB}(\Lambda)}\,,\quad
    u_I(\Lambda)=-\mu_I + \mu_B\frac{\lambda_{BI}(\Lambda)}{\lambda_{BB}(\Lambda)}\,,
\end{align}
where $\mu_B>0$, and $\mu_I/\mu_B < \lambda_{BI}(\Lambda)/\lambda_{BB}(\Lambda)$.

To connect the flow to known physical observables, we impose that the couplings $\lambda_{BB}$ and $\lambda_{BI}$ in vacuum ($\mu_B=\mu_I=0$) correspond to the known two-body $T$-matrices at the physical limit $k=0$ (see details in Ref.~\cite{floerchinger_functional_2008}). With this, the initial conditions for $\lambda_{BB}$ and $\lambda_{BI}$ depend on the boson-boson and boson-impurity scattering lengths $a_{BB}$ and $a_{BI}$, respectively. For the  optimized 
regulator~(\ref{sec:RP;eq:Rlitim}), we have the same initial conditions as those for the repulsive Bose-Bose mixtures studied in Ref.~\cite{isaule_functional_2021}. They are
\begin{equation}
   \lambda_\alpha(\Lambda)=
\begin{cases}
\dfrac{2\pi/m_\alpha}{1-2\gamma_E-\ln(a_\alpha^2\Lambda^2/4)} & :d=2\\[10pt]
\left(\dfrac{m_\alpha}{2\pi a_\alpha}-\dfrac{m_\alpha}{3\pi^2}\Lambda\right)^{-1} & :d=3
\end{cases}\,,
\label{sec:RP;sub:IC;eq:lambdaIC}
\end{equation}
where $\alpha=BB,BI$, and $\gamma_E\approx 0.577$ is the Euler-Mascheroni constant. The 
reduced masses are $m_{BB}=m_B/2$ and $m_{BI}=m_r=m_Bm_I/(m_B+m_I)$. For purely repulsive potentials, the scattering lengths provide a lower bound to the potential ranges. Thus, a contact potential approximation becomes invalid for momenta larger than the inverse scattering length \cite{isaule_thermodynamics_2020,isaule_functional_2021}.
The flow must therefore be restricted to $\Lambda<\min(a_{BI}^{-1},a_{BB}^{-1})$. Nevertheless, we stress that because the interactions are renormalized by Eq.~(\ref{sec:RP;sub:IC;eq:lambdaIC}), as long as $\Lambda\gg k_h$, the results are independent of the choice of $\Lambda$. For more details see Refs.~\cite{floerchinger_functional_2008,isaule_thermodynamics_2020}.

The initial conditions completely define the RG flow in terms of the physical inputs $\mu_B$, $a_{BB}$, and $a_{BI}$, and the self-consistently determined $\mu_I$. We then follow the RG flow by solving the flow equations of all the couplings.
Note that we choose values of $a_{BB}$ and $\mu_B$ which give the desired physical density of the bath for $k\to 0$~\cite{floerchinger_functional_2008}.
Examples of flows are given in App.~\ref{app:RP}.

\subsection{Propagator and polaron energy}
\label{sec:RP;sub:G}

As explained in Sec.~\ref{sec:model}, the propagator of the FRG equation is given by $\MG_k=(\MGamma_k^{(2)}+\MR_k)^{-1}$. In momentum space $q=(\omega,\Q)$, 
the inverse propagator for ansatz (\ref{sec:RP;eq:Gamma}) reads
\begin{equation}
    \MG^{-1}_k(q)=\begin{pmatrix}
    \MG^{-1}_{k,B}(q) & \bm{0}\\
    \bm{0} & \MG^{-1}_{k,I}(q)
    \end{pmatrix}\,,
\end{equation}
where
\begin{equation}
    \MG^{-1}_{k,B}(q)=\begin{pmatrix}
    E_{1,k}(\Q;\rho_B)+V_B\omega^2 & S_B\omega\\
    -S_B\omega &  E_{2,k}(\Q;\rho_B)+V_B\omega^2 
    \end{pmatrix}\,,
  \label{sec:RP;sub:G;eq:GinvB}
\end{equation}
is the inverse propagator of the Bose gas, with
\begin{align}
     E_{1,k}(\Q;\rho_B)=&E_{2,k}(\Q;\rho_B)+2\rho_BU_B''(\rho_B)\,,\label{sec:RP;sub:G;eq:E1}\\
    E_{2,k}(\Q;\rho_B)=&Z_B\frac{\Q^2}{2m_B}+U_B'(\rho_B)+R_{k,B}(\Q)\,,\label{sec:RP;sub:G;eq:E2}
\end{align}
where the primes in $U'$ and $U''$ indicate derivatives with respect to $\rho_B$, whereas
\begin{equation}
    \MG^{-1}_{k,I}(q)=\begin{pmatrix}
    E_{I,k}(\Q;\rho_B)+iS_I\omega & 0\\
   0 &  E_{I,k}(\Q;\rho_B)-iS_I\omega
    \end{pmatrix}\,,
  \label{sec:RP;sub:G;eq:GinvI}
\end{equation}
is the inverse propagator of the impurity, with
\begin{equation}
     E_{I,k}(\Q;\rho_B)=Z_I\frac{\Q^2}{2m_I}+\partial_{\rho_I}U(\rho_B,\rho_I)+R_{k,I}(\Q)\,.
     \label{sec:RP;sub:G;eq:EI}
\end{equation}
Note that we introduced real orthogonal fields 
$\psi_B=(\psi_{B,1}(x)+i\psi_{B,2}(x))/\sqrt{2}$, and evaluated all the fields at their background 
values $\psi_{B,1}=\sqrt{2\rho_B}\delta(q)$ and $\psi_{B,2}=\psi_I=\phi=0$~\cite{floerchinger_functional_2008}.

The polaron energy $\mu_I$ corresponds to the energy needed to add an impurity to the medium. In the ground state, the Green's function of the impurity $\MG_I$ 
(or, analogously, the spectral function) has a pole at $\mu_I$~\cite{rath_field-theoretical_2013}. 
In our FRG formalism, we find the ground state energy by determining the 
energy $\mu_I$ that gives $\det(\MG^{-1}_{k,I}(0))=0$ for $k\to 0$~\cite{schmidt_excitation_2011}.

From Eq.~(\ref{sec:RP;sub:G;eq:GinvI}), by taking $\det(\MG^{-1}_{k,I})=0$ at the minimum $\rho_B=\rho_0$, we find the pole
\begin{equation}
    q_0^*(\Q)=E_{I,k}(\Q)/S_I\,,
    \label{sec:RP;sub:G;eq:q0}
\end{equation}
where $E_{I,k}$ is defined in Eq.~(\ref{sec:RP;sub:G;eq:EI}). At zero momentum, $q_0^*(\bm{0})=u_I/S_I$. Therefore, the physical polaron energy $\mu_I^*$ corresponds to the choice of $\mu_I$ which gives $q_0^*(\bm{0})\to 0$ for $k\to 0$. Values of $\mu_I$ that do not fulfill this condition are not physical. An analogous condition is imposed to find the ground state of the Fermi polaron~\cite{schmidt_excitation_2011}, and binding energies in few-boson problems~\cite{floerchinger_efimov_2011}. We note that because in the DE we follow the flow at zero momentum $q=0$ (see App.~\ref{app:RP}), we can not study the poles at finite momenta in the current work.

\subsection{Results}
\label{sec:RP;sub:results}

Following the approach sketched above, here we present results for the polaron energy for a range of boson-impurity scattering lengths $a_{BI}$. This scattering length can be tuned experimentally through Feshbach resonances~\cite{jorgensen_observation_2016}. We present results in both two and three dimensions and compare them with known results to check the robustness of our approach.

\begin{figure}[t]
\centering
\includegraphics[scale=0.70]{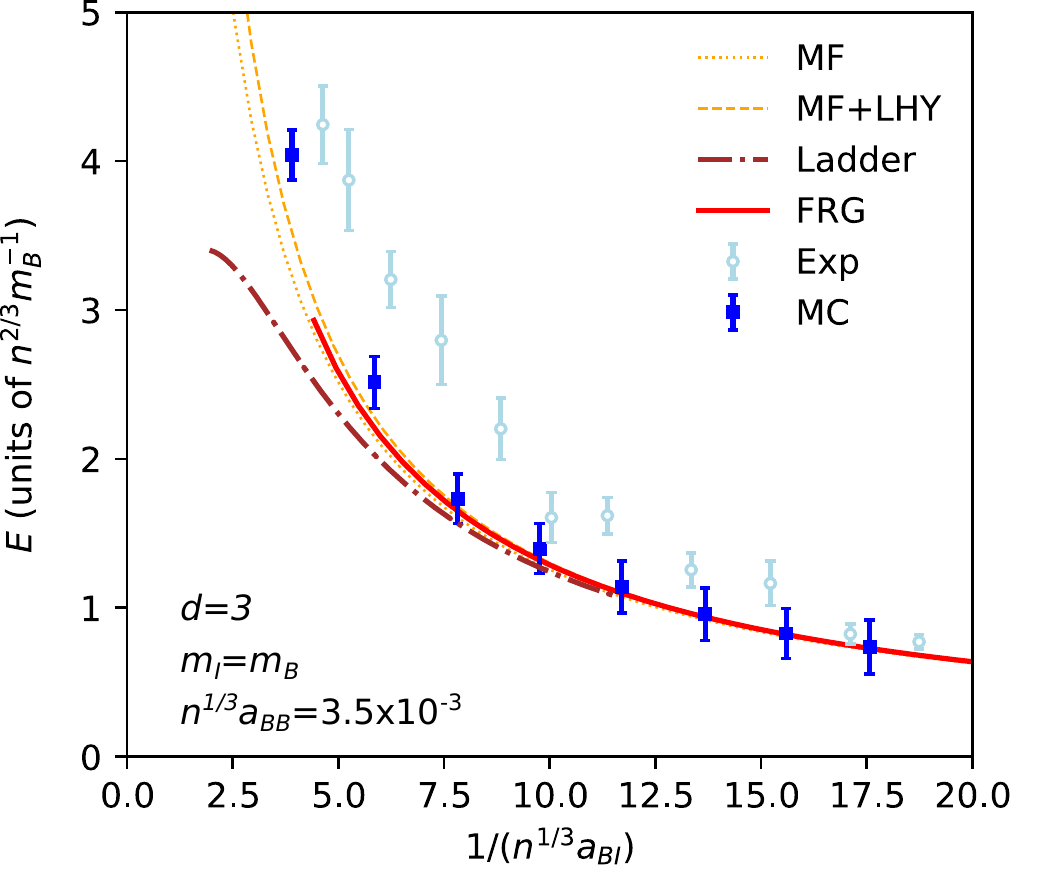}
\caption{Polaron energy $E$ of the repulsive branch in three dimensions as a function of $(n^{1/3}a_{BI})^{-1}$. The solid red line corresponds to FRG calculations. 
The dash-dotted brown line corresponds to ladder calculations from Ref.~\cite{camacho-guardian_landau_2018}.
The thin orange lines correspond to the perturbative 
solution~(\ref{sec:RP;sub:results;eq:Epert3D}) at the MF level (dotted) and with the 
LHY-type correction (dashed). The light blue open circles are experimental data from Ref.~\cite{pena_ardila_analyzing_2019},
and the blue squares are MC simulations from Ref.~\cite{pena_ardila_analyzing_2019}. In all cases, 
$m_B=m_I$, and $n^{1/3}a_{BB}=3.5\times 10^{-3}$.}
\label{sec:RP;sub:results;fig:3D_3o5Em3_1}
\end{figure}

First, we show results in three dimensions in Fig.~\ref{sec:RP;sub:results;fig:3D_3o5Em3_1}. We employ parameters that simulate the conditions of the Aarhus experiment~\cite{jorgensen_observation_2016}, and scattering lengths $a_{BI}>a_{BB}$, so the effect of the boson-impurity interaction is important. 
We compare with MC simulations and experimental data from Ref.~\cite{pena_ardila_analyzing_2019}, ladder calculations from Ref.~\cite{camacho-guardian_landau_2018},  and with the perturbative solution~\cite{viverit_ground-state_2002,christensen_quasiparticle_2015}
\begin{equation}
    E=\frac{2\pi a_{BI}n}{m_R}\left[1+\frac{24}{3\sqrt{\pi}}\frac{m_R}{m_I}\sqrt{n a_{BB}^3}\frac{a_{BI}}{a_{BB}}I(\gamma)\right]\,,
    \label{sec:RP;sub:results;eq:Epert3D}
\end{equation}
where $n$ is the density of the bosonic bath, $\gamma=m_B/m_I$, and
\begin{equation*}
    I(\gamma)=\frac{1+\gamma}{\gamma}\int_0^\infty dk\left[1-\frac{(1+\gamma)k^2}{\sqrt{1+k^2}(\sqrt{1+k^2}+\gamma k)}\right]\,,
\end{equation*}
which for equal masses takes the value $I(1)=8/3$. The first term in 
Eq.~(\ref{sec:RP;sub:results;eq:Epert3D}) corresponds to the mean field (MF) 
solution, whereas the second term is a Lee-Huang-Yang (LHY) type correction.

\begin{figure}[t]
\centering
\includegraphics[scale=0.70]{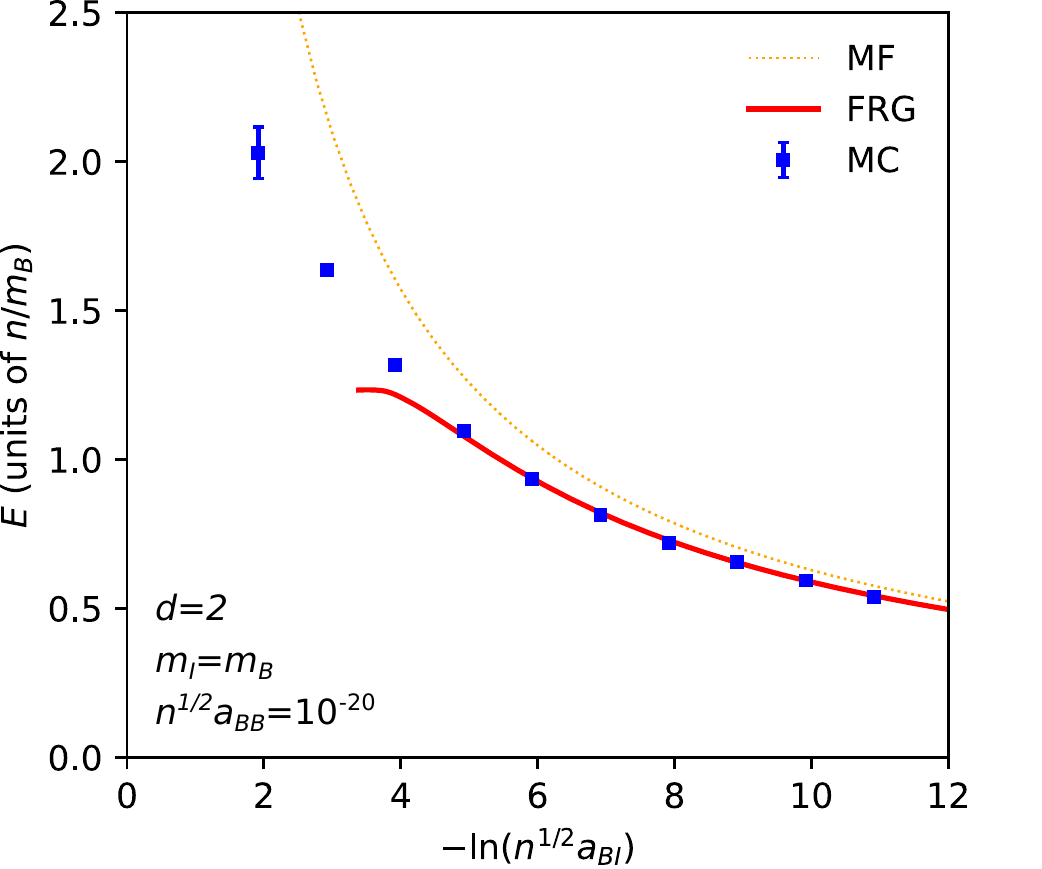}
\caption{Polaron energy $E$ of the repulsive branch in two dimensions as a function of $-\ln(n^{1/2}a_{BI})$. The solid red line corresponds to FRG calculations. The thin dotted orange line corresponds to the MF solution~(\ref{sec:RP;sub:results;eq:Epert2D}). The blue squares are MC simulations from Ref.~\cite{pena_ardila_strong_2020}. In all cases, 
$m_B=m_I$, and $n^{1/2}a_{BB}=10^{-20}$.}
\label{sec:RP;sub:results;fig:2D_1m20_1}
\end{figure}

We obtain good agreement between our results and both the MC simulations and perturbative solutions. This agreement is in line with previous FRG results for repulsive Bose-Bose mixtures~\cite{isaule_functional_2021}. We stress that the MC simulations include fluctuations at all orders, and thus they are a good benchmark for our calculations.
We restrict our 
calculations to $(n^{1/3}a_{BI})^{-1}\gtrsim 4 $, 
because for stronger boson-impurity 
interactions we have that $k_h \gtrsim a_{BI}^{-1}$, and thus, we can not choose a sufficiently large value for the initial scale $\Lambda$ [see discussion after Eq.~\eqref{sec:RP;sub:IC;eq:lambdaIC}].

We show an analogous calculation in two dimensions in Fig.~\ref{sec:RP;sub:results;fig:2D_1m20_1}. We employ conditions that have been achieved experimentally in two-dimensional traps with $^{87}$Rb atoms~\cite{desbuquois_superfluid_2012,hung_observation_2011}. 
We compare with MC simulations from Ref.~\cite{pena_ardila_strong_2020} and with the MF solution~\cite{pena_ardila_strong_2020}
\begin{equation}
    E=\frac{\pi n/m_R}{|\ln(n^{1/2}a_{BI})|}\,.
    \label{sec:RP;sub:results;eq:Epert2D}
\end{equation}
As in three dimensions, we obtain excellent agreement with MC simulations for $-\ln(n^{1/2}a_{BI})\gtrsim 4$. Furthermore, the FRG results show an important improvement over the perturbative solution. This is expected, as the FRG 
has proved to give a good description of 
two-dimensional gases~\cite{floerchinger_superfluid_2009,rancon_universal_2012}. 
In contrast, perturbative results are less reliable in two dimensions because of the enhanced effect of fluctuations. 
For stronger interactions $(-\ln(n^{1/2}a_{BI})\lesssim 4)$ the FRG 
calculations become unreliable due to the breakdown of the initial conditions.

We have checked that we obtain similarly good descriptions of the polaron energy also for other choices of gas parameters and masses $m_B \approx m_I$  in both two and three dimensions. We also provide an estimate of the effective mass in App.~\ref{app:meff}.

\section{Attractive branch}
\label{sec:AP}

We switch now to the description of the quasiparticle excitation which is present at negative energies, the so-called attractive polaron.

\begin{figure}[t!]
\centering
\includegraphics[scale=0.35]{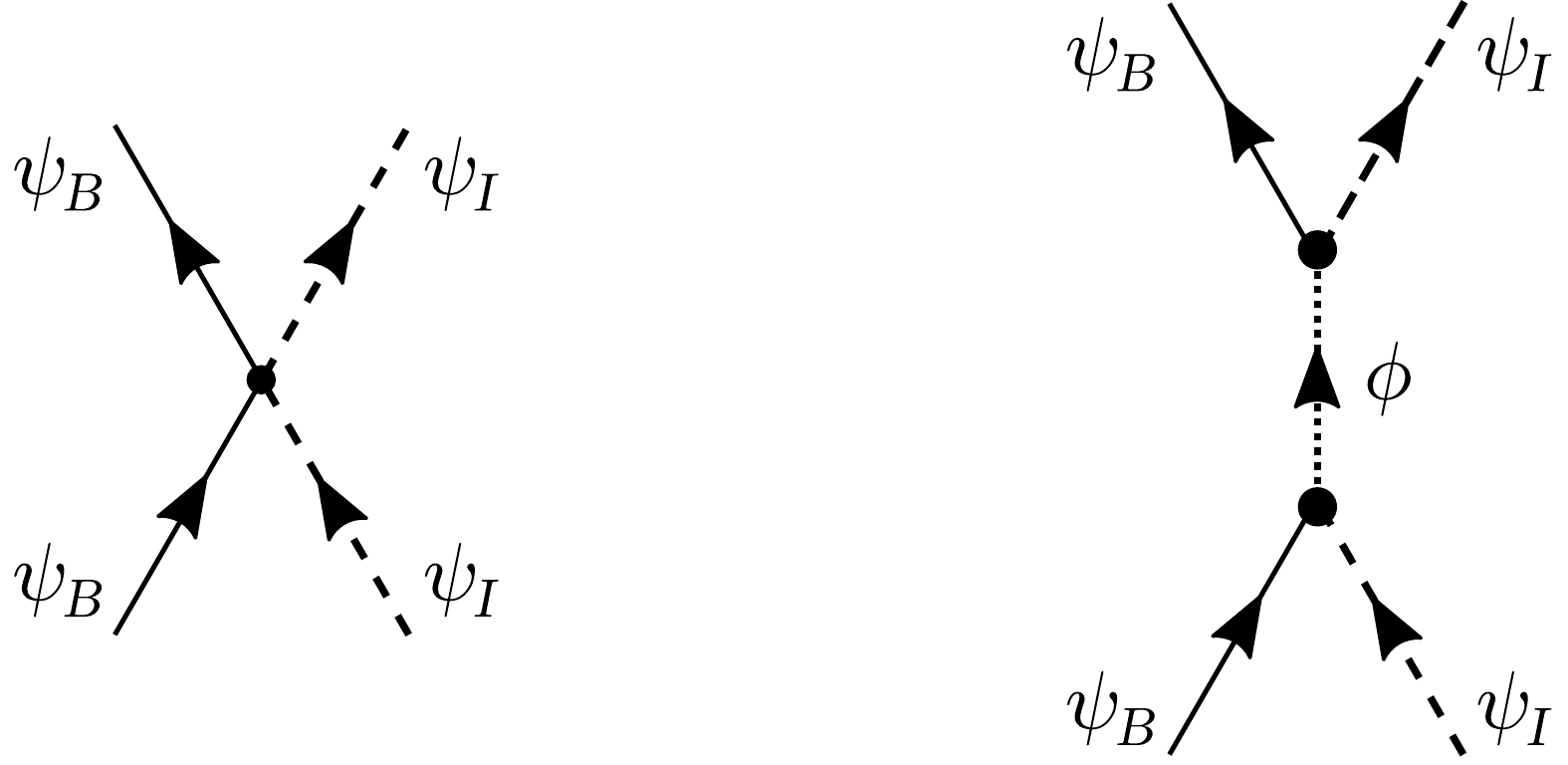}
\caption{Tree-level diagram for the scattering between a boson and an impurity before (left) and after (right) the introduction of the auxiliary dimer fields. Solid, dashed, and dotted lines denote bosons, impurities, and dimers, respectively.}
\label{sec:AP;fig:HS}
\end{figure}

Because we now deal with attractive interactions, we have to consider the formation of bound states. In particular, the Bose polaron shows a polaron-to-molecule crossover~\cite{rath_field-theoretical_2013}. Two-body bound states appear as poles in the four-point vertices $\MGamma^{(4)}$. However, in a straightforward application of the DE [see Eq.~(\ref{sec:RP;eq:Gamma})], all the terms in the expansion are regular, and thus they do not account for bound states~\cite{floerchinger_efimov_2011}. We can circumvent this by introducing dimer fields $\phi\sim\psi_B\psi_I$ to mediate the boson-impurity interaction via a Hubbard-Stratonovich transformation~\cite{schmidt_excitation_2011}. Analogous transformations are used in FRG studies of Fermi gases~\cite{diehl_universality_2006,diehl_functional_2007} and few atoms~\cite{floerchinger_efimov_2011}.
In the context of Feshbach resonances, the fields $\psi_B$ and $\psi_I$ represent atoms in the open channel, whereas the field $\phi$ represents dimers in the closed channel~\cite{diehl_universality_2006}. The resulting action takes the form a two-channel model~\cite{rath_field-theoretical_2013}
\begin{align}
    \mathcal{S}[\VPHI]=&\int_x \bigg[\psid_B\left(\partial_\tau-\frac{\nabla^2}{2m_B}-\mu_B\right)\psi_B\nonumber\\
    &+\psid_I\left(\partial_\tau-\frac{\nabla^2}{2m_I}-\mu_I\right)\psi_I\nonumber\\
    &+\phid\left(\partial_\tau-\frac{\nabla^2}{2m_\phi}+\nu_\phi\right)\phi+\frac{g_{BB}}{2}(\psid_B\psi_B)^2\nonumber\\
    &+h\left(\phid\psi_B\psi_I+\phi\psid_B\psid_I\right)\bigg]\,,
    \label{sec:AP;eq:SHS}
\end{align}
where $m_\phi=m_B+m_I$ is the mass of a closed-channel dimer, $\nu_\phi$ is the dimer detuning, and $h$ is the \emph{Feshbach coupling}. The Hubbard-Stratonovich transformation is illustrated 
in Fig.~\ref{sec:AP;fig:HS}. 

We work in the \emph{broad resonance limit} where $h,\nu_\phi\to\infty$ but 
$h^2/\nu_\phi$ is kept constant~\cite{diehl_universality_2006}. In this limit, by integrating out the dimer fields in Eq.~(\ref{sec:AP;eq:SHS}) we recover the original one-channel model~(\ref{sec:model;eq:S}), and thus both 
equations are equivalent. Therefore, we stress that in this work Eq.~(\ref{sec:AP;eq:SHS}) physically describes a 
one-channel model where $\phi$ simply acts as an auxiliary field.

\begin{figure}[t!]
\centering
\includegraphics[scale=0.35]{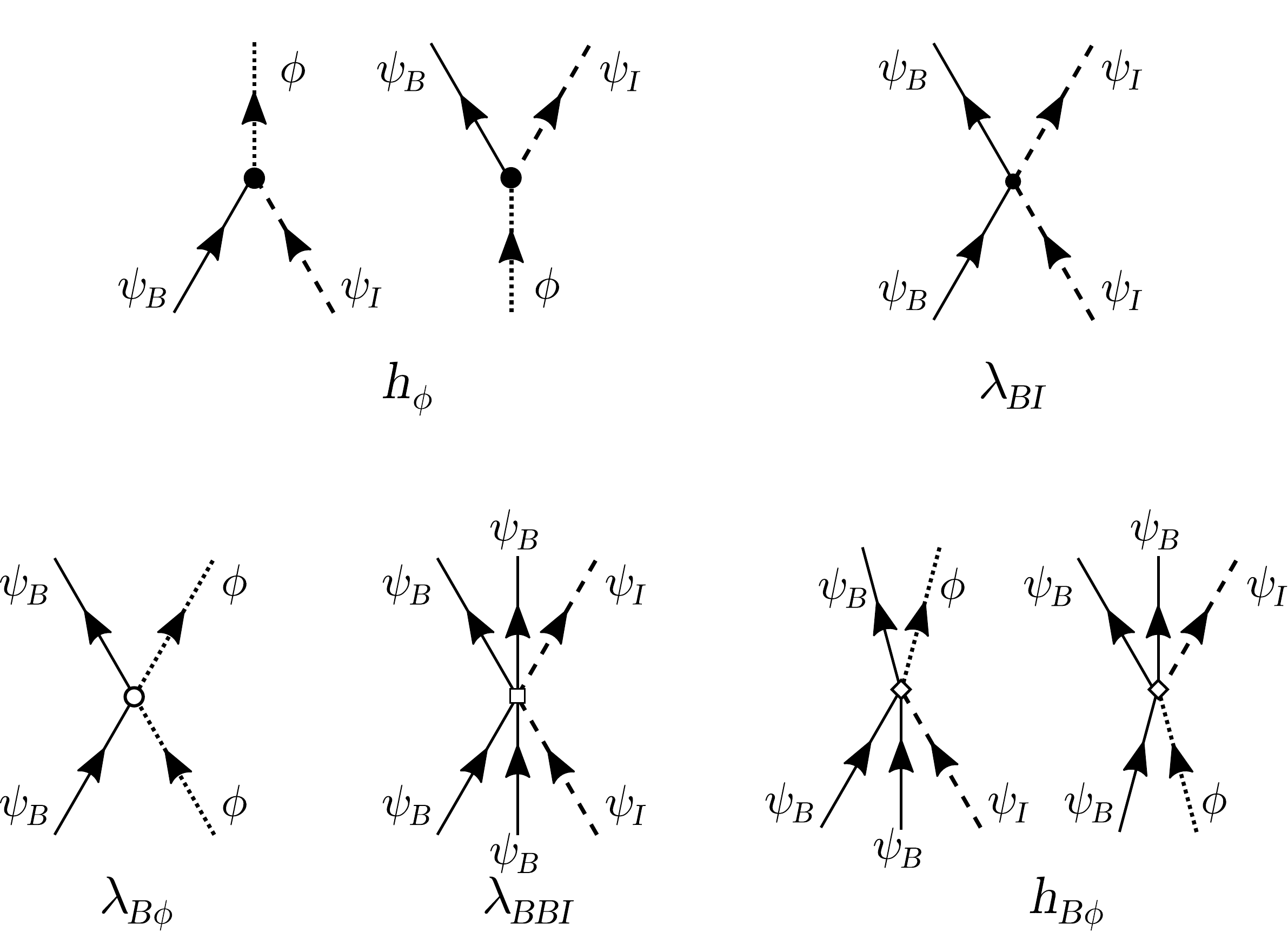}
\caption{Two- (top row) and three-body (bottom row) vertices associated with the 
interaction between the bosons and the impurity. Solid lines represent the bosons, 
dashed lines represent the impurity, and the dotted lines represent the dimer field.}
\label{sec:AP;fig:vertices}
\end{figure}

Based on action (\ref{sec:AP;eq:SHS}), we propose the following ansatz for the effective action in the attractive branch:
\begin{align}
    \Gamma_k[\PHI]=&\int_x \bigg[\psid_B\left(S_B\partial_\tau-\frac{Z_B}{2m_B}\nabla^2-V_B\partial^2_\tau\right)\psi_B\nonumber\\
    &+\psid_I\left(S_I\partial_\tau-\frac{Z_I}{2m_I}\nabla^2+U_I(\rho_B)\right)\psi_I\nonumber\\
    &+\phid\left(S_\phi\partial_\tau-\frac{Z_\phi}{2m_\phi}\nabla^2+U_\phi(\rho_B)\right)\phi\nonumber\\
    &+H_\phi(\rho_B)\left(\phid\psi_B\psi_I+\phi\psid_B\psid_I\right)+U_B(\rho_B)\bigg]\,,
\label{sec:AP;eq:ansatz}
\end{align}
where $\rho_B=\psid_B\psi_B$. Our ansatz is similar to those used for the Fermi polaron~\cite{schmidt_excitation_2011,von_milczewski_functional_2021} and, as the repulsive branch, is valid in only two and three dimensions. Note that the dimer fields become dynamical, with 
flowing renormalization factors $S_\phi$ and $Z_\phi$. We expand the boson effective 
potential as
\begin{equation}
    U_B=-P+\frac{\lambda_{BB}}{2}(\rho_B-\rho_0)^2\,,
\end{equation}
analogously to Eq.~(\ref{sec:RP;eq:U}). As in the repulsive branch, $\rho_0$ and $\rho_s=Z_B\rho_0$ at $k=0$ correspond to the physical condensate and superfluid densities of the bosonic medium, respectively. The rest of the functions contain the interactions between the bosonic medium and the impurity. Because three-body correlations are important in the attractive branch, we expand these up to 
three-body couplings 
\begin{align}
    U_I&=u_I+\lambda_{BI}(\rho-\rho_0)+\frac{\lambda_{BBI}}{2}(\rho-\rho_0)^2\,,\label{sec:AP;eq:UI}\\
    U_\phi&=u_\phi+\lambda_{B\phi}(\rho-\rho_0)\,,\label{sec:AP;eq:Uphi}\\
    H_\phi&=h_\phi+h_{B\phi}(\rho-\rho_0)\,.\label{sec:AP;eq:Hphi}
\end{align}
Here, $h_\phi$ and $\lambda_{BI}$ correspond to two-body boson-impurity vertices, 
whereas $\lambda_{B\phi}$, $\lambda_{BBI}$, and $h_{B\phi}$ correspond to three-body vertices. 
These vertices are illustrated in Fig.~\ref{sec:AP;fig:vertices}.

The FRG framework enables us to include the effect of three-body correlations by allowing the three-body couplings to flow. In the following, we want to test the relevance of three-body effects. To switch those off, it is enough to fix the corresponding couplings at zero for all $k$.

Similarly to the repulsive branch, all the renormalization factors ($Z_a$, $S_a$, $V_B$) and the couplings in $U_B$, $U_I$, $U_\phi$, and $H_\phi$, as well as $\rho_0$, flow with $k$.
We also employ the optimized regulator~(\ref{sec:RP;eq:Rlitim}) for all the fields $a=B,I,\phi$. We provide the flow equations in App.~\ref{app:AP}, and we examine the initial conditions here below.

\subsection{Initial conditions of the RG flow}
\label{sec:AP;sub:IC}

Similarly to the repulsive branch, the RG flow is started at a high scale 
$k=\Lambda$ much larger than the healing scale of the bath $k_h=(2m_B\mu_B)^{1/2}$. 
By imposing that $\Gamma_\Lambda=\mathcal{S}$, we obtain
\begin{align}
    S_B(\Lambda)=Z_B(\Lambda)= S_I(\Lambda)=Z_I(\Lambda)=1\, \nonumber\\ V_B(\Lambda)=0\,,\quad\rho_0(\Lambda)=\frac{\mu_B}{\lambda_{BB}(\Lambda)}\,,\quad u_I=-\mu_I\,,\label{sec:AP;sub:IC;eq:IC}
\end{align}
where $\mu_B>0$ and $\mu_I<0$. Note that, in contrast to the repulsive branch, 
the impurity energy $\mu_I$ is negative by construction.

The couplings $\lambda_{BB}$, $u_\phi$, $Z_\phi$, and $S_\phi$ are renormalized 
in vacuum so they can be connected to physical  
scattering~\cite{floerchinger_functional_2008,diehl_functional_2007}. The initial 
condition for $\lambda_{BB}$ is given by Eq.~(\ref{sec:RP;sub:IC;eq:lambdaIC}). For 
the boson-impurity interaction, we consider the boson-impurity scattering 
length $a_{BI}$ and effective range $r_0$ as physical inputs.
The effective range is necessary to have a well defined three-body sector with attractive interactions in three dimensions. Otherwise, the UV is not well defined since the infinite tower of Efimov trimers which appears in this case lacks a reference scale~\cite{yoshida_universality_2018}. 
At low collision energies, the 
boson-impurity $T$ matrix takes the form~\cite{khuri_low-energy_2009,galea_fermions_2017} 
\begin{equation}
    T_{BI}=
\begin{cases}
\dfrac{2\pi/m_r}{\ln(-4/p_R^2 a_{BI}^2)-2\gamma_E-\frac{\pi r_0^2}{4}p_R^2+i\pi} & :d=2\\[10pt]
\dfrac{2\pi/m_r}{a_{BI}^{-1}-\frac{r_0}{2} p_R^2+i p_R} & :d=3
\end{cases}\,,
\end{equation}
where
\begin{equation}
    p_R=-\sqrt{2m_r\left(p_0-\frac{\vP^2}{2m_\phi}+\mu_\phi\right)}\,,
\end{equation}
is the relative momentum, with $\mu_\phi=\mu_B+\mu_I$ and $p_0=i\omega_p$. At the physical limit $k=0$ in vacuum, we impose that (see Ref.~\cite{birse_functional_2008} for details)
\begin{equation}
    \frac{h_\phi^2}{\Pi_\phi(p_0,\vP)}\bigg|_{k=0}=-T_{BI}\,,
\end{equation}
where $\Pi_\phi$ is the full dimer self energy. $\Pi_\phi$ is related to 
the couplings in ansatz~(\ref{sec:AP;eq:ansatz}) through $u_\phi=\Pi_\phi(0,\bm{0})$, and
\begin{align}
    Z_\phi=&2m_\phi\frac{\partial}{\partial \vP^2}\Pi_\phi\Big|_{p_0=0,\vP=\bm{0}}\,,\\
    S_\phi=&-\frac{\partial}{\partial p_0}\Pi_\phi\Big|_{p_0=0,\vP=\bm{0}}\,.
\end{align}
We obtain the following initial conditions~\cite{birse_functional_2008}:
\begin{equation}
   \frac{u_\phi}{h^2_\phi}\bigg|_\Lambda=
\begin{cases}
\dfrac{m_r}{2\pi}\left(\ln(a_{BI}^2\Lambda^2/4)+2\gamma_E-1\right) & :d=2\\[10pt]
\dfrac{2m_r}{3\pi^2}\Lambda-\dfrac{m_r}{2\pi a_{BB}} & :d=3
\end{cases}\,,
\label{sec:flow;sub:IC;eq:uphi}
\end{equation}
and
\begin{equation}
   \frac{Z_\phi}{h^2_\phi}\bigg|_\Lambda=\frac{S_\phi}{h^2_\phi}\bigg|_\Lambda=
\begin{cases}
\dfrac{m_r^2}{\pi}\left(\dfrac{2}{\Lambda^2}-\dfrac{\pi}{4}r_0^2\right)& :d=2\\[10pt]
\dfrac{m_r^2}{\pi^2}\left(\dfrac{8}{3\Lambda}-\dfrac{\pi}{2}r_0\right) & :d=3
\end{cases}\,.
\label{sec:flow;sub:IC;eq:Zphi}
\end{equation}
Note that $u_\phi$ and $h_\phi$ are chosen freely as long as they 
satisfy Eq.~(\ref{sec:flow;sub:IC;eq:uphi}). 

Since we work in the broad resonance limit, we naturally have $r_0>0$~\cite{chin_feshbach_2010,Tanzi2018}. In three dimensions, a positive effective range ensures that the dimer fields become non-dynamical in the UV with $Z_\phi,S_\phi\to 0$, so our ansatz describes a one-channel model. For details on the FRG for narrow resonances, see Ref.~\cite{diehl_universality_2006}.

The scattering length $a_{BB}$ sets a lower bound for the range of the boson-boson interaction, whereas $r_0$ sets the range of the attractive boson-impurity interaction. Analogously to the repulsive branch, we must restrict the flow to momenta smaller than $a^{-1}_{BB}$ and $r^{-1}_0$ in order for our approximation of contact potentials to be valid. Therefore, the initial scale has to satisfy $\Lambda < \min(a_{BB}^{-1},r_0^{-1})$. We stress again that because of the renormalizations (\ref{sec:flow;sub:IC;eq:uphi}) and (\ref{sec:flow;sub:IC;eq:Zphi}), the results are independent of the choice of $\Lambda$ as long as $\Lambda\gg k_h$. 

In contrast to purely repulsive potentials, for attractive potentials the scattering length can be tuned independently of the range. Therefore, in the attractive branch we can choose $a_{BI}$ freely. This enables us to study the regime of strong boson-impurity coupling, including the unitary limit $a_{BI}\to\infty$
in three dimensions~\cite{diehl_functional_2007}. In contrast, in the repulsive branch the initial scale $\Lambda$ is heavily restricted by $a_{BI}$ [see discussion after Eq.~(\ref{sec:RP;sub:IC;eq:lambdaIC})].

The couplings not mentioned so far are not present in the microscopic theory (\ref{sec:AP;eq:SHS}), so their values at $k=\Lambda$ are zero. In particular, the three-body couplings are zero in the UV, and they are generated only as $k$ is lowered. To capture the effect of three-body correlations at high scales, we must start the RG flow 
at a high scale near the range of the interactions: $\Lambda \approx \min(a_{BB}^{-1},r_0^{-1})$. 

\subsection{Propagator and polaron energy}
\label{sec:AP;sub:G}

In the attractive branch, the inverse propagator for ansatz (\ref{sec:AP;eq:ansatz}) reads
\begin{equation}
    \MG^{-1}_k(q)=\begin{pmatrix}
    \MG^{-1}_{k,B}(q) & \bm{0}\\
    \bm{0} & \MG^{-1}_{k,I\phi}(q)
    \end{pmatrix}\,,
\end{equation}
where $G_{k,B}^{-1}$ is given in Eq.~(\ref{sec:RP;sub:G;eq:GinvB}), and
\begin{widetext}
\begin{equation}
    \MG^{-1}_{k,I\phi}(q)=\begin{pmatrix}
    E_{I,k}(\Q;\rho_B)+iS_I\omega & 0 & \rho_B^{1/2}H_\phi(\rho_B) & 0\\
   0 &  E_{I,k}(\Q;\rho_B)-iS_I\omega & 0 & \rho_B^{1/2}H_\phi(\rho_B) \\
   \rho_B^{1/2}H_\phi(\rho_B) & 0 & E_{\phi,k}(\Q;\rho_B)+iS_\phi\omega & 0 \\
   0 & \rho_B^{1/2}H_\phi(\rho_B) & 0 & E_{\phi,k}(\Q;\rho_B)-iS_\phi\omega
    \end{pmatrix}\,,
    \label{sec:AP;sub:G;eq:GinvIphi}
\end{equation}
\end{widetext}
is the impurity-dimer inverse propagator, with
\begin{align}
     E_{I,k}(\Q;\rho_B)&=Z_I\frac{\Q^2}{2m_I}+U_I(\rho_B)+R_{k,I}(\Q)\,,\label{sec:AP;sub:G;eq:EI}\\
     E_{\phi,k}(\Q;\rho_B)&=Z_\phi\frac{\Q^2}{2m_\phi}+U_\phi(\rho_B)+R_{k,\phi}(\Q)\,.\label{sec:AP;sub:G;eq:Ephi}
\end{align}
In contrast to the problem of an impurity in a Fermi bath, whose ground state features a sharp transition between a polaron and a dressed dimer,
the Bose polaron problem shows a smooth polaron-to-molecule crossover. We can understand this 
from Eq.~(\ref{sec:AP;sub:G;eq:GinvIphi}), where if $\rho_0>0$ we can not 
separate $\MG^{-1}_{k,I\phi}$ into independent impurity and dimer propagators. 
Therefore, the impurity and dimer propagators are hybridized, and we can not 
identify a polaron or a molecule phase (for more details, we refer
to Ref.~\cite{rath_field-theoretical_2013}).

To find the ground state energy $\mu_I$, we search for the pole of the Green's function $\MG_{I\phi}$, as we did for the repulsive branch in Sec.~\ref{sec:RP;sub:G}. 
By taking $\det(\MG^{-1}_{k,I\phi})=0$ at the minimum $\rho_B=\rho_0$, we find 
two poles,
\begin{widetext}
\begin{equation}
    q_{0,\pm}^*(\Q)=\frac{1}{2}\Bigg[\frac{E_{I,k}(\Q)}{S_I}+\frac{E_{\phi,k}(\Q)}{S_\phi}
    \pm\sqrt{\left(\frac{E_{I,k}(\Q)}{S_I}+\frac{E_{\phi,k}(\Q)}{S_\phi}\right)^2-\frac{4}{S_I S_\phi}(E_{I,k}(\Q)E_{\phi,k}(\Q)-h_\phi^2\rho_0)}\Bigg]\,,
    \label{sec:AP;sub:G;eq:q0pm}
\end{equation}
\end{widetext}
where $E_I$ and $E_\phi$ are defined in Eqs.~(\ref{sec:AP;sub:G;eq:EI}) 
and (\ref{sec:AP;sub:G;eq:Ephi}). As with the repulsive branch, we identify 
the choice of $\mu_I$ that gives $q_{0,\pm}^*(\bm{0})=0$ for $k\to 0$ as the 
energy of the polaron. Similarly, choices of $\mu_I$ that do not fulfill this condition are not physical.

We find that $q_{0,+}^*(\bm{0})$ and $q_{0,-}^*(\bm{0})$ go to zero at the same impurity energy 
$\mu_I$, and thus, there is one ground state energy for each combination of 
interaction strengths. This can change at finite temperatures, where the 
spectrum can split into more than one quasiparticle~\cite{guenther_bose_2018,field_fate_2020}.
We provide details of the flows in App.~\ref{app:AP}.

\subsection{Results}
\label{sec:AP;sub:results}

In the following, we evaluate the polaron energy in both two and three dimensions, and we compare it with available analytical and experimental results. To quantify the effect of three-body correlations, we present curves which include only two-body correlations (2B) and both two- and three-body correlations (2B+3B).

\subsubsection{Three dimensions}
\label{sec:AP;sub:results;sub:3D}

\begin{figure}
\centering
\includegraphics[scale=0.70]{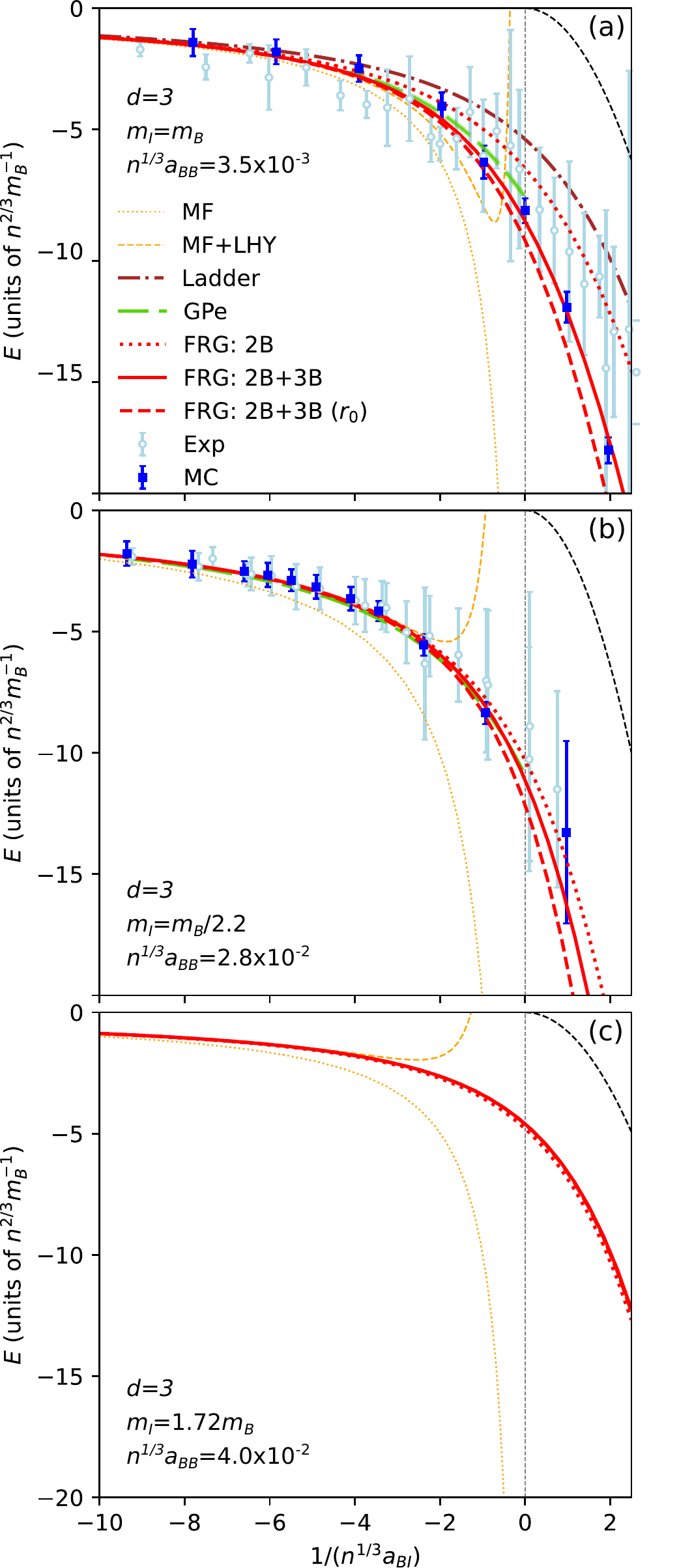}
\caption{Polaron energy in three dimensions as a function of $(n^{1/3}a_{BI})^{-1}$. 
The red lines correspond to FRG calculations with only 
2B interactions (dotted), 2B+3B interactions with $r_0=0$ (solid), and 2B+3B 
interactions with $r_0=3,1.5,0.6a_{BB}$ (dashed) for (a), (b), and (c), respectively. 
The thin orange lines show the perturbative 
solution~(\ref{sec:RP;sub:results;eq:Epert3D}) at the MF level (dotted), and with the 
first correction (dashed). The green dashed lines are GPe calculations from Ref.~\cite{guenther_mobile_2021} [in panel (b) the green line is underneath the solid red line]. The dash-dotted brown line shows ladder calculations from Ref.~\cite{camacho-guardian_landau_2018}, the blue squares are MC simulations 
from Ref.~\cite{pena_ardila_analyzing_2019}, and the light blue circles are experimental data from Refs.~\cite{jorgensen_observation_2016,pena_ardila_analyzing_2019} 
(a) and \cite{hu_bose_2016} (b). The dashed black lines show the binding 
energy in vacuum $\epsilon_b=-(2m_R a_{BI}^2)^{-1}\Theta(a_{BI})$.}
\label{sec:AP;sub:results;sub:3D;fig:A_E_cBI}
\end{figure}

The attractive branch of the three-dimensional polaron at zero temperature was previously investigated in various works. To test the robustness of our FRG calculations, we compare our results with MC simulations, solutions of the Gross-Pitaevskii equation (GPe), ladder calculations, and experimental data.

Fig.~\ref{sec:AP;sub:results;sub:3D;fig:A_E_cBI} shows the polaron energy for three choices of masses and gas parameters. The panels (a), (b) and (c) simulate, respectively, the conditions of the Aarhus~\cite{jorgensen_observation_2016}, JILA~\cite{hu_bose_2016}, and MIT~\cite{yan_bose_2020}  
experiments. We present calculations with only two-body
interactions and $r_0=0$ (because for the interactions we investigated the 
two-body sector is only weakly sensitive to $r_0$), and with two- and three-body interactions 
with both $r_0=0$ and $r_0>0$. 
We use the effective ranges $r_0$ computed in Ref.~\cite{guenther_mobile_2021}.

In panels (a) and (b) we compare our results with experimental data, GPe calculations~\cite{guenther_mobile_2021}, and MC simulations~\cite{pena_ardila_analyzing_2019}. In 
addition, we compare with the perturbative solution~(\ref{sec:RP;sub:results;eq:Epert3D}), 
where $I(2.2)\approx 1.78$ and $I(1/1.72)\approx 1.99$. Additionally, in panel (a) we compare with ladder calculations from Ref.~\cite{camacho-guardian_landau_2018}, which give an upper bound for the energy. We observe a noticeable effect 
of three-body correlations in panels (a) and (b), as well of the effective range. In contrast, in panel (c) three-body effects are not as important. This is in agreement with previous studies that showed that three-body effects are more important for 
lighter impurities and at 
lower bath densities~\cite{sun_visualizing_2017,yoshida_universality_2018}.

In the weakly interacting regime $(n^{1/3}a_{BI})^{-1}\lesssim -4$, 
our FRG calculations recover the expected result from perturbation theory~(\ref{sec:RP;sub:results;eq:Epert3D}). For stronger interactions, $(n^{1/3}a_{BI})^{-1}\gtrsim -4$, our FRG is in very good agreeement with both MC and GPe results.
We obtain the best agreement with MC by including three-body correlations with $r_0=0$. Nevertheless, also our calculations with finite effective range are also in reasonable agreement with experiment.

\begin{figure}
\centering
\includegraphics[scale=0.70]{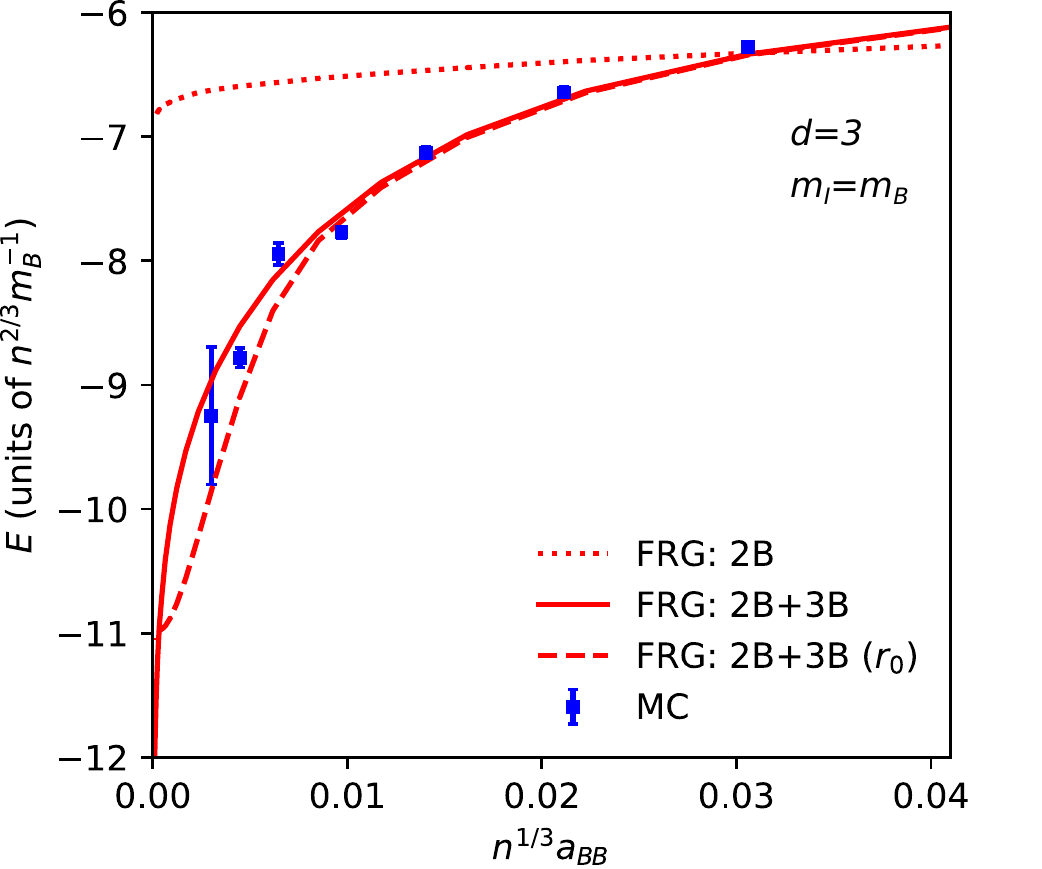}
\caption{Polaron energy in three dimensions at unitarity $a_{BI}\to\infty$ for $m_B=m_I$ as a function of the gas parameter $n^{1/3}a_{BB}$. The red lines correspond to FRG calculations with only 2B interactions (dotted), 2B+3B interactions with $r_0=0$ (solid), and 2B+3B interaction with $r_0=2.2\times 10^{-3}/(m_B\mu_B)^{1/2}>0$. The blue circles are MC simulations from Ref.~\cite{pena_ardila_impurity_2015}.
}
\label{sec:AP;sub:results;sub:3D;fig:A_E_U_cBB}
\end{figure}

To examine the effect of the gas parameter, in Fig.~\ref{sec:AP;sub:results;sub:3D;fig:A_E_U_cBB} we show the polaron energy as a function of $n^{1/3}a_{BB}$ at unitarity, where three-body correlations are important.
We find that the effect of three-body correlations depends strongly on the gas parameter, in agreement with previous studies~\cite{yoshida_universality_2018}. As we approach the vacuum limit $n^{1/3}a_{BB}\to 0$, three-body correlations become much more important, significantly decreasing the polaron energy. In particular, 
we obtain good agreement with MC simulations from Ref.~\cite{pena_ardila_impurity_2015} by considering three-body effects. This confirms both the importance of three-body physics and also the robustness of our calculations.

At low gas parameters, the polaron energy is more sensitive to the effective range. In particular, with three-body interactions, the polaron energy goes to a finite value with a finite effective range, whereas we are not able to find a bound with $r_0=0$. As explained in detail in Ref.~\cite{yoshida_universality_2018}, 
with $r_0=0$ there are infinite Efimov trimers in vacuum, and so the energy diverges at the vacuum limit. In contrast, for $r_0>0$ there is a well defined deepest Efimov state, and therefore the energy saturates to a finite value. Finally, let us mention that the limit of very low gas parameters is extremely delicate \cite{yoshida_universality_2018,guenther_mobile_2021, massignan_universal_2021,levinsen_quantum_2021}. 
Indeed, in the limit $a_{BB}\rightarrow 0$ the bath becomes infinitely compressible, and thus, multi-body correlations have increasing importance. 
To study such a regime with FRG, one would need to include further higher-order couplings in the ansatz.

\subsubsection{Two dimensions}
\label{sec:AP;sub:results;sub:2D}

The two-dimensional Bose polaron was studied in detail only recently, with MC simulations in Ref.~\cite{pena_ardila_strong_2020}. Here, we provide results for various conditions achieved in current experiments~\cite{desbuquois_superfluid_2012,hung_observation_2011}.

\begin{figure}[t!]
\centering
\includegraphics[scale=0.70]{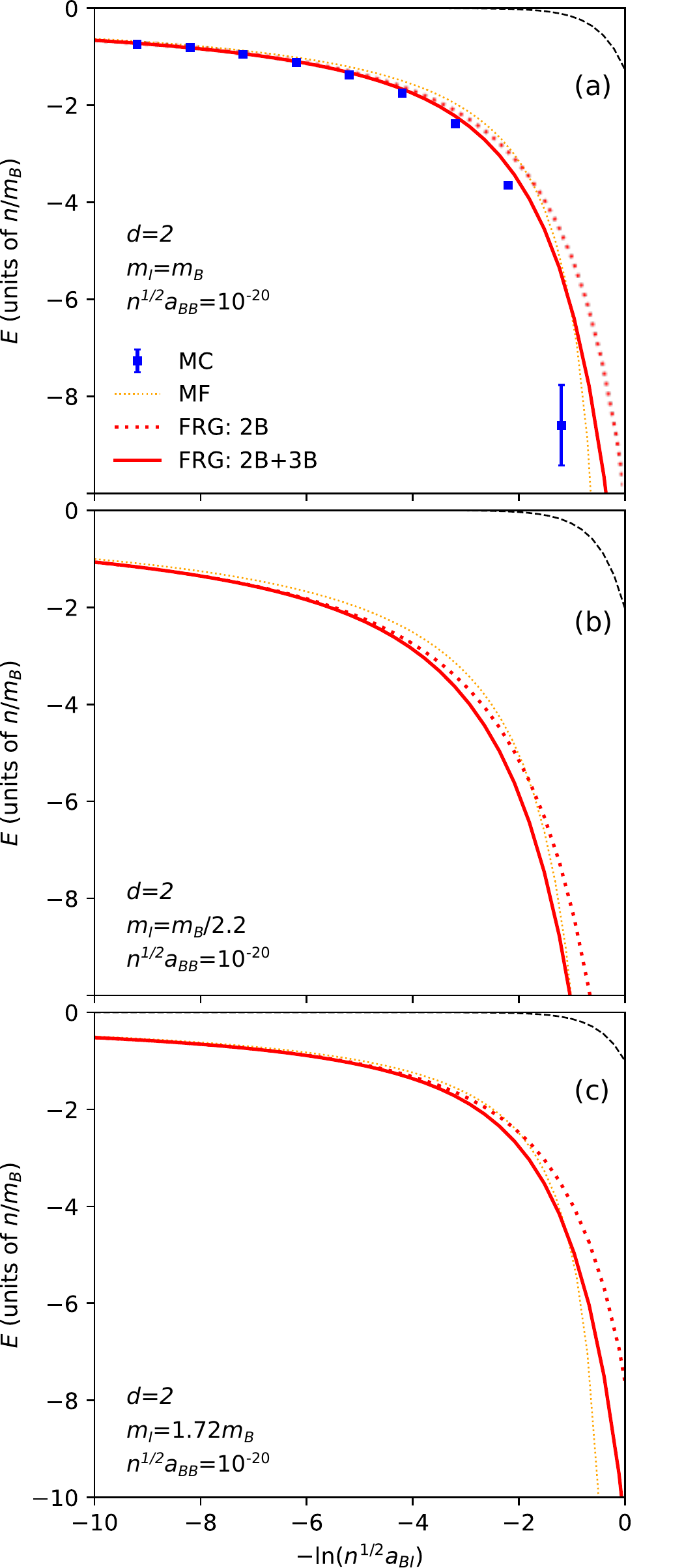}
\caption{Polaron energy in two dimensions as a function of $-\ln(n^{1/2}a_{BI})$. Masses and gas parameters are given within the figures. The red lines correspond to FRG calculations with only 2B interactions (dotted), and 2B+3B interactions with $r_0=0$ (solid). The thin orange gray lines show the MF solution~(\ref{sec:RP;sub:results;eq:Epert2D}). The blue squares are MC simulations from Ref.~\cite{pena_ardila_strong_2020}. The dashed black lines are the boson-impurity binding energy $\epsilon_b=-2/(m_re^{2\gamma_E}a^2_{BI})$.}
\label{sec:AP;sub:results;sub:2D;fig:A_E_cBI}
\end{figure}

Fig.~\ref{sec:AP;sub:results;sub:2D;fig:A_E_cBI} shows results for three different 
combinations of masses and the gas parameter $n^{1/2}a_{BI}=10^{-20}$. All 
the calculations use $r_0=0$. We find that our calculations are insensitive 
to reasonably chosen effective ranges. This result is not unexpected. In two dimensions, 
there are no Efimov trimers in vacuum, with only two three-body bound states~\cite{hammer_universal_2004}.

In panel (a) we compare our results with MC simulations from Ref.~\cite{pena_ardila_strong_2020}. Additionally, we compare them with the perturbative solution~(\ref{sec:RP;sub:results;eq:Epert2D}).
In all cases, we obtain a noticeable effect of three-body correlations. In particular, in panel (a) we obtain better agreement with the MC simulations by considering three-body effects. However, we do not obtain agreement as good as in three dimensions. This could be an effect either of the derivative expansion or of not considering higher-order couplings. We stress that because fluctuations are enhanced in low dimensions, it is expected that our approximation is less robust than in three dimensions.

\begin{figure}[t!]
\centering
\includegraphics[scale=0.70]{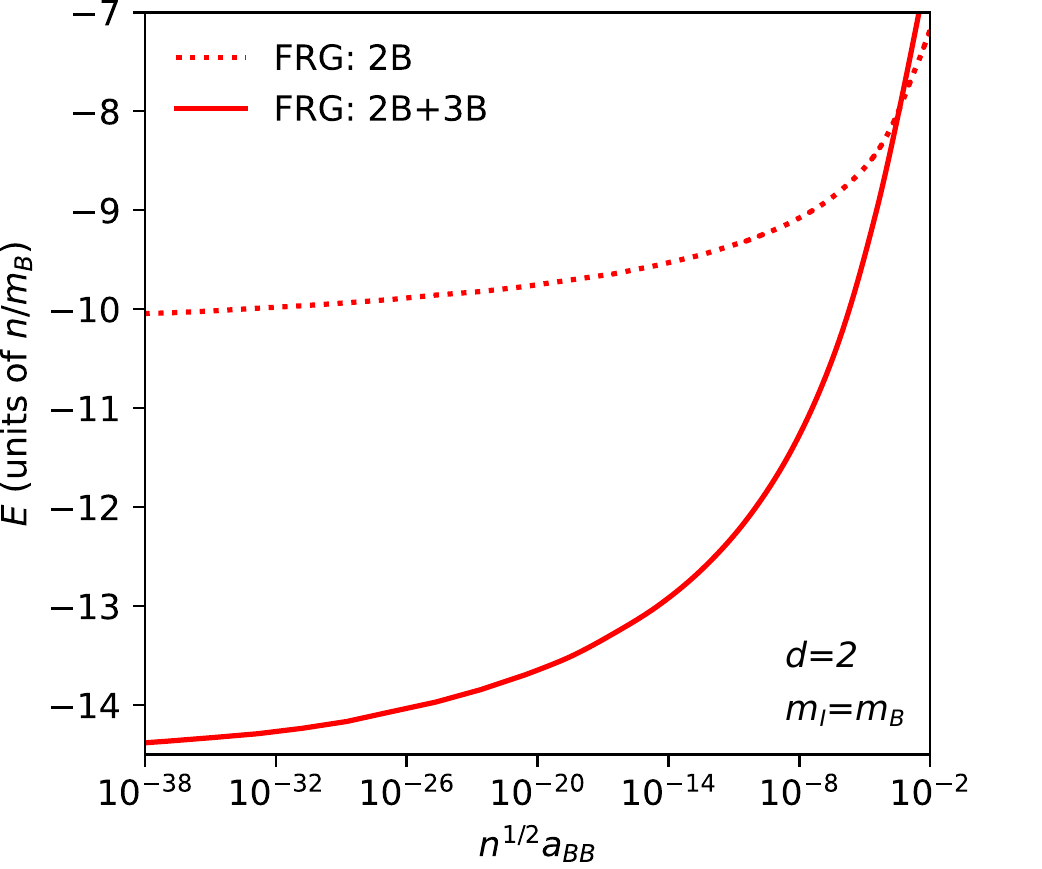}
\caption{Polaron energy in two dimensions at $\ln(n^{1/2}a_{BI})=0$ for $m_B=m_I$ as a function of the gas parameter $n^{1/2}a_{BB}$. The red lines show FRG calculations with only 2B interactions (dotted) and 2B+3B interactions with $r_0=0$ (solid).}
\label{sec:AP;sub:results;sub:2D;fig:A_E_0_cBB}
\end{figure}

To study the effect of the density of the medium, in Fig.~\ref{sec:AP;sub:results;sub:2D;fig:A_E_0_cBB} we show the polaron energy in the strong coupling regime as a function of the bath density. We show results for $\ln(n^{1/2}a_{BI})=0$, which can not be described by the perturbative solution~(\ref{sec:RP;sub:results;eq:Epert2D}). We explore a wide range of gas parameters. We note that recent experiments have produced two-dimensional bosonic gases with gas parameters as high as $n^{1/2}a_{BB}\approx 10^{-9} - 10^{-4}$ 
without many losses~\cite{ha_strongly_2013,christodoulou_observation_2021}.

As in three dimensions, we observe an important effect of three-body correlations. However, the energy seems to converge to a finite value for the vacuum gas limit with and without three-body effects, even with zero effective range. This is expected. As mentioned, in two dimensions there are only two well defined tree-body bound states in vacuum~\cite{hammer_universal_2004} instead of infinite Efimov trimers. Nevertheless, we do not reach the value of the binding energy of the deepest trimer in vacuum $E\approx 16.5 \epsilon_b$, where $\epsilon_b=-2/(m_re^{2\gamma_E}a^2_{BI})$. In Fig.~\ref{sec:AP;sub:results;sub:2D;fig:A_E_0_cBB}, this corresponds to $E m_B/n\approx -21$. This is probably due to our truncation of the derivative expansion, which does not describe few-body physics accurately~\cite{floerchinger_efimov_2011}. We expect that the inclusion of further couplings will improve the convergence.

\section{Conclusions}
\label{sec:concl}

In this work, we studied the Bose polaron at zero temperature in two and three dimensions with the FRG. We approximated the effective action by means of a derivative expansion, which enabled us to find the ground state energies of the polaron by following the flow of the scale-dependent poles of the impurity's propagator. 

We studied both the repulsive and attractive branches of the Bose polaron. In the attractive branch, we introduced dimer fields via a Hubbard-Stratonovich transformation to mediate the boson-impurity interaction. This enabled us to access the regime of strong coupling easily. In addition, in the attractive branch, we added the effect of three-body correlations by considering up to three-body couplings in the derivative expansion.

We obtained polaron energies in good agreement with state-of-the-art theoretical and experimental results.
In particular, in the attractive branch, we obtained the best agreement by adding three-body effects. Overall, we showed that the FRG can successfully describe the regime of strong coupling in both two and three dimensions.

Throughout this manuscript, we focused on cases where $0.5\lesssim m_I/m_B\lesssim 2$. The reason is twofold. On one side, for heavy impurities, homogeneous fields might not provide a good description of static particles. On the other side, for light impurities, the attractive branch 
is strongly influenced by Efimov trimers~\cite{sun_visualizing_2017}, which require careful treatment, beyond the scope of the current work.

Having demonstrated that the Bose polaron can be successfully described with the FRG, there are several extensions of this work that we plan to explore in the future. First, we intend to consider the full momentum dependence of the flowing couplings by employing a vertex expansion~\cite{berges_non-perturbative_2002} in order to give a more robust description of the Bose polaron. This will enable us to obtain the full Green's function, which is not accessible within the derivative expansion, and to study dynamical properties, as well as decay rates. Furthermore, the account for momentum dependent vertices is necessary to accurately capture few-body physics, including the onset of Efimov states, which are beautifully captured with the FRG as periodic cycles in the RG flow~\cite{floerchinger_efimov_2011}. We also plan to explore the effect of four- and more-body correlations by adding higher-order couplings.

On top of the current works on Bose and Fermi polarons, the FRG could be used to study impurities in other scenarios. A small finite number of impurities is a natural extension of this work, and systems with a small population of impurities can be studied as quantum mixtures with large population imbalances~\cite{boettcher_phase_2015,von_milczewski_functional_2021}. Furthermore, polarons in optical lattices could be naturally studied by employing the lattice implementation of the FRG~\cite{machado_local_2010}, which has proved very successful in describing strongly-correlated lattice gases~\cite{rancon_nonperturbative_2011}. 
Using impurities to probe topological excitations is, presently, a topic of great interest \cite{MunozDeLasHeras2020,Grass2020,Pimenov2021}, and the FRG may be a good tool to address them at strong coupling. 
Finally, polarons at finite temperatures can be easily studied by using the Matsubara formalism. Particularly interesting would be to examine the impact of the BKT transition on the properties of two-dimensional Bose polarons.

\begin{acknowledgments}
We thank L.~Peña Ardila for providing the MC data from Refs.~\cite{pena_ardila_impurity_2015,pena_ardila_analyzing_2019,pena_ardila_strong_2020}, M.~G.~Skou for the experimental data from Ref.~\cite{pena_ardila_analyzing_2019}, N.~Guenther for the GPe data from Ref.~\cite{guenther_bose_2018}, and A. Camacho-Guardian for the ladder calculations~\cite{camacho-guardian_landau_2018}. We also thank G. Bruun for the careful reading of our manuscript. 
This  research  has  been  supported by MINECO (Spain) Grants No.~FIS2017-87034-P and FIS2017-84114-C2-1-P,  and by the National Science Foundation under Grant No. NSF PHY-1748958.  We also acknowledge financial support from Secretaria d’Universitats i Recerca del Departament d’Empresa i Coneixement de la Generalitat de Catalunya, co-funded by the European Union Regional Development Fund within the ERDF Operational Program of  Catalunya (project QuantumCat, ref.~001-P-001644).
 
\end{acknowledgments}

\appendix

\section{RG flow of the repulsive branch}
\label{app:RP}

\subsection{Flow equations}
\label{app:RP;sub:eqs}

In the DE all the couplings are momentum-independent, $f(q)=f$, and their flows are simply obtained by differentiating the Wetterich equation ~(\ref{sec:model;eq:WettEq}). Because there is no feedback of the impurity onto the medium, the flow equations for the bosonic couplings ($\rho_0$, $\lambda_{BB}$, $Z_B$, $S_B$ and $V_B$) are identical to those of a one-component Bose gas. Thus, we refer to Ref.~\cite{floerchinger_functional_2008} for details. Also, note that it is not necessary to follow the flow of the $k$-dependent pressure $P$, as it does not affect the flow of the rest of the couplings. 

The flows of the couplings associated with the impurity are dictated by
\begin{align}
    \partial_k u_I =&\partial_k\Gamma^{(2)}_{k,I^\dag I}\Big|_{\rho_0,p=0}\,,\\
    \partial_k \lambda_{BI} =&\frac{\partial}{\partial\rho_B}(\partial_k\Gamma^{(2)}_{k,I^\dag I})\Big|_{\rho_0,p=0}\,,\\
    \partial_k Z_I=&2m_I\frac{\partial}{\partial\vP^2}(\partial_k\Gamma^{(2)}_{k,I^\dag I})\Big|_{\rho_0,p=0}\,,\\
    \partial_k S_I=&i\frac{\partial}{\partial\omega_p}(\partial_k\Gamma^{(2)}_{k,I^\dag I})\Big|_{\rho_0,p=0}\,,
\end{align}
where $\partial_k\Gamma_k$ is obtained from the Wetterich equation~(\ref{sec:model;eq:WettEq}), $p=(\omega_p,\vP)$ is an external momentum which is taken to zero after differentiating, and the two-point function is defined as
\begin{equation}
    \Gamma^{(2)}_{k,I^\dag I}=\frac{\delta^2\Gamma}{\delta\psid_I\delta\psi_I}\,.
    \label{app:RP;sub:eqs;eq:Gamma2II}
\end{equation}
We note that we take $p=0$ because in a DE all the couplings are momentum-independent, and thus we follow the flow at zero momentum~\cite{dupuis_nonperturbative_2021}. Studies at finite momenta can be implemented within a vertex expansion, where the ansatz for $\Gamma$ is proposed in terms of momentum-dependent vertices instead of simple derivatives~\cite{berges_non-perturbative_2002}. However, to solve the RG flow in a vertex expansion, we usually need to perform sophisticated calculations, such as with the BMW approximation~\cite{blaizot_non-perturbative_2005}, which are beyond the scope of this work. 

\begin{figure}[t!]
\centering
\includegraphics[scale=0.35]{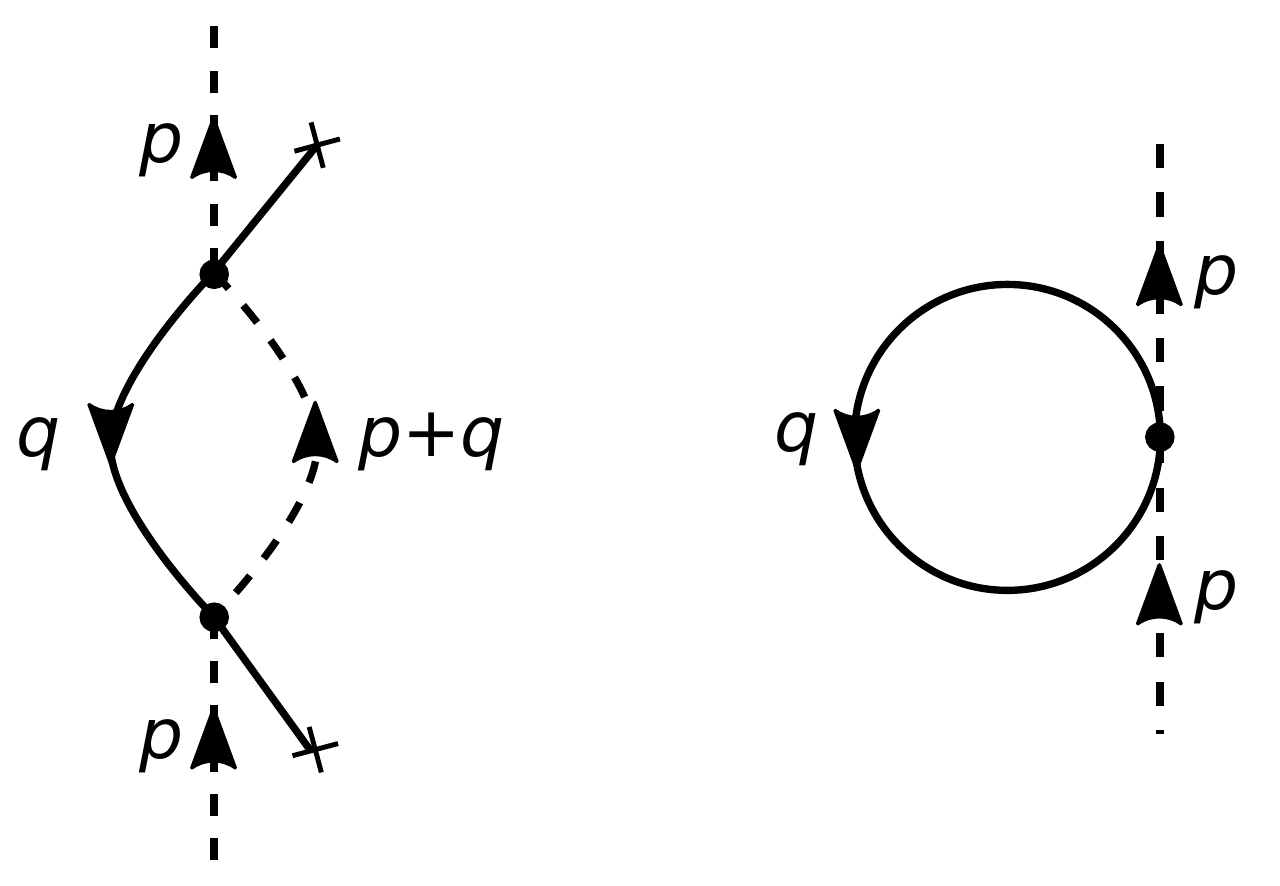}
\caption{Diagrams that contribute to the flow of $\Gamma^{(2)}_{k,I^\dag I}$. Solid and dashed lines denote bosons and impurities, respectively. The cross denotes a boson field evaluated at its background value $\langle\psi_B\rangle=\sqrt{\rho_B}$.}
\label{app:RP;sub:eqs;fig:Gamma2II}
\end{figure}

Because we follow the flow at the minimum $\rho_0$, we evaluate at $\rho_B=\rho_0$ after taking the derivatives. The diagrams that contribute to the flow of $\Gamma^{(2)}_{k,I^\dag I}$ are shown in Fig.~\ref{app:RP;sub:eqs;fig:Gamma2II}. They give the following expression
\begin{align}
 \partial_k \Gamma^{(2)}_{k,I^\dag I}=&\tilde{\partial}_k\int_q\bigg[4\rho_B\lambda^2_{BI}\frac{(E_{2,k}(\Q)+V_B\omega^2)E_{I,k}(\Q+\vP)}{\det_B(q)\det_I(q+p)}\nonumber\\
 &-\frac{\lambda_{BI}(E_{1,k}(\Q)+E_{2,k}(\Q)+2V_B\omega^2)}{2\det_B(q)}\bigg]\,,
\end{align}
where $\tilde{\partial}_k$ is a $k$-derivative that acts on only the regulators, $E_1$, $E_2$ and $E_I$ are given in Eqs.~(\ref{sec:RP;sub:G;eq:E1},\ref{sec:RP;sub:G;eq:E2}) and (\ref{sec:RP;sub:G;eq:EI}), and
\begin{align}
    \text{det}_B(q)=&S_B^2\omega^2+(E_{1,k}(\Q)+V_B\omega^2)(E_{2,k}(\Q)+V_B\omega^2)\,,\label{app:RP;sub:eqs;eq:detB}\\
    \text{det}_I(q)=&S_I^2\omega^2+E_{I,k}(\Q)^2\,.
\end{align}
Note that $E_1$, $E_2$ and $E_I$ still depend on $\rho_B$, which is taken to $\rho_B=\rho_0$ only after taking all the derivatives. Finally, the integral over internal momentum $q=(\omega,\Q)$ is defined as
\begin{equation}
    \int_q=\int_{-\infty}^\infty\frac{d\omega}{2\pi}\int\frac{d^d\Q}{(2\pi)^d}\,.
\end{equation}

\subsection{Examples of flows}
\label{app:RP;sub:flows}

\begin{figure}[t!]
\centering
\includegraphics[scale=0.70]{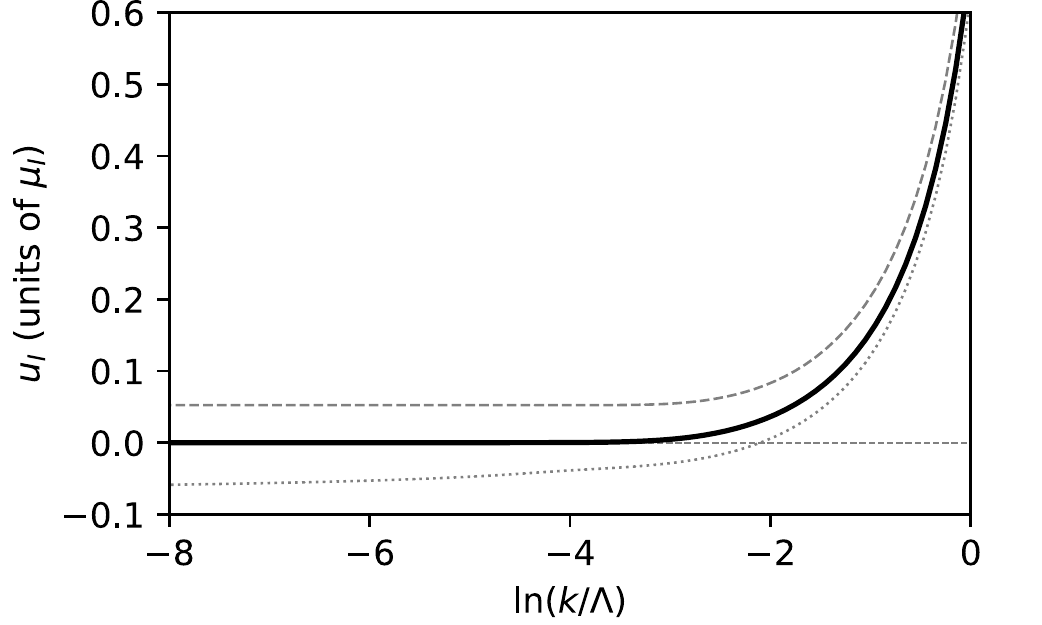}
\caption{Flow of $u_I$ in three dimensions in the repulsive branch for $n^{1/3}a_{BB}=3.5\times 10^{-3}$ at $(n^{1/3}a_{BI})^{-1}=10$. The solid line corresponds to the flow obtained for the ground state energy $\mu_I^*$, the thin dashed line to a flow obtained with $\mu_I<\mu_I^*$, and the thin dotted to a flow obtained with $\mu_I>\mu_I^*$.}
\label{app:RP;sub:flows;fig:uI}
\end{figure}

Fig.~(\ref{app:RP;sub:flows;fig:uI}) shows the flow of $u_I$ for different $\mu_I$ for the chosen parameters for the bosonic medium and the boson-impurity scattering length. At the physical polaron energy $\mu_I^*$, the coupling $u_I$ flows to zero (solid black line), giving a vanishing $q_0^*(\bm{0})$. In contrast, for $\mu_I<\mu_I^*$ the coupling $u_I$ saturates to finite values greater than zero, whereas for $\mu_I<\mu_I^*$ it goes to negative values. Thus, values of $\mu_I\neq\mu_I^*$ do not correspond to physical energies of the impurity.

This behavior is found for any combination of parameters in both two and three dimensions. 

\section{RG flow of the attractive branch}
\label{app:AP}

\subsection{Flow equations and $k$-dependent dimer fields}
\label{app:AP;sub:eqs}

The strategy is similar to that used for the repulsive branch in App.~\ref{app:RP}, where the flow of the momentum-independent couplings is obtained by differentiating the Wetterich equation. The flow of the bosonic couplings ($\rho_0$, $\lambda_{BB}$, $Z_B$, $S_B$, and $V_B$) is given by those of a one-component Bose gas~\cite{floerchinger_functional_2008}, whereas the flow of the couplings in Eqs.~(\ref{sec:AP;eq:UI}-\ref{sec:AP;eq:Hphi}) can be extracted from the flow of the two point functions
\begin{align}
    \partial_k U_I &= \partial_k\Gamma^{(2)}_{I^\dag I}\Big|_{p=0}\,, \\
    \partial_k U_\phi &= \partial_k\Gamma^{(2)}_{\phi^\dag \phi}\Big|_{p=0}\,, \\
    \partial_k H_\phi &= \partial_k\Gamma^{(2)}_{I^\dag \phi}\Big|_{p=0}\,,
\end{align}
where the derivatives in $\Gamma^{(2)}$ are defined in the same way as in Eq.~(\ref{app:RP;sub:eqs;eq:Gamma2II}). We provide their explicit expression in the next subsection. The specific flow of the different couplings within $U_I$, $U_\phi$, and $H_\phi$ are obtained by taking $\rho_B$ derivatives and then evaluating at $\rho_B=\rho_0$. Also, as with the repulsive branch, we follow the flow at zero momentum, and thus we evaluate at $p=0$.

\begin{figure}[t!]
\centering
\includegraphics[scale=0.35]{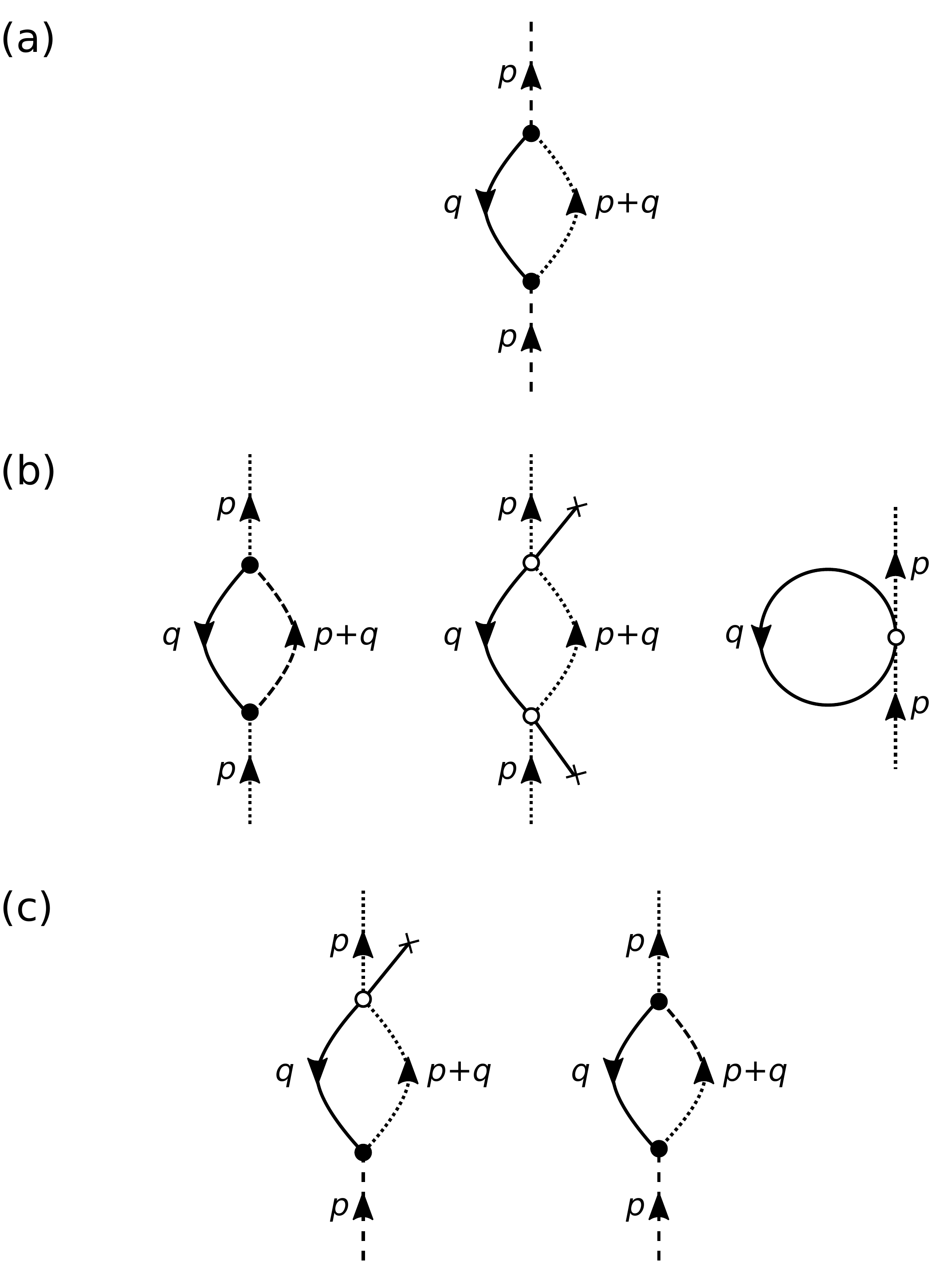}
\caption{Diagrams that contribute to the flow of $\Gamma^{(2)}_{k,I^\dag I}$ (a), $\Gamma^{(2)}_{k,\phi^\dag \phi}$ (b), and $\Gamma^{(2)}_{k,\phi^\dag I}$ (c). Solid, dashed, and dotted lines denote bosons, impurities, and dimers, respectively. The crosses denotes boson fields evaluated at their background value $\langle\psi_B\rangle=\sqrt{\rho_B}$.}
\label{app:AP;sub:Gamma2;fig:Gamma2}
\end{figure}

To simplify the flow equations, we introduce $k$-dependent dimer fields $\phi_k$ to eliminate the flow of some of the couplings. If we introduce $k$-dependent fields, the Wetterich equation becomes~\cite{berges_non-perturbative_2002}
\begin{equation}
\partial_k \Gamma_k=\frac{1}{2}\tr\left[(\MGamma_k^{(2)}+\MR_k)^{-1}\partial_k \MR_k\right]+\frac{\delta\Gamma_k}{\delta\PHI}\cdot\partial_k\PHI_k\,.
\label{app:AP;eq:WettEq}
\end{equation}
We choose the $k$-dependent fields so that they eliminate the flow of the couplings $\lambda_{BI}$, $\lambda_{BBI}$, and $h_{B\phi}$. Similar eliminations have been used in FRG studies of few bosons~\cite{schmidt_renormalization-group_2010,jaramillo_avila_universal_2013,jaramillo_avila_four-boson_2015} and Fermi gases~\cite{floerchinger_particle-hole_2008}. This elimination means that these couplings will not flow with $k$ and will remain at zero during the entire flow. We use $k$-dependent dimer fields defined as
\begin{align}
    \partial_k \phi_k=&f_{2B}(k)\psi_B\psi_I+f_{3B}(k)(\rho_B-\rho_0)\psi_B\psi_I\nonumber\\
    &+g_{3B}(k)(\rho-\rho_0)\phi_k\,,\\
    \partial_k \phid_k=&f_{2B}(k)\psid_B\psid_I+f_{3B}(k)(\rho_B-\rho_0)\psid_B\psid_I\nonumber\\
    &+g_{3B}(k)(\rho-\rho_0)\phid_k\,,
\end{align}
where the functions $f_{2B}$, $f_{3B}$, and $g_{3B}$ are chosen to eliminate the flows of $\lambda_{BI}$, $\lambda_{BBI}$ and $h_{B\phi}$. Following Eq.~(\ref{app:AP;eq:WettEq}), the flow of these couplings is dictated by
\begin{align}
    \partial_k \lambda_{BI}=&\partial_k \lambda_{BI}\Big|_\phi+2f_{2B}h_\phi+2f_{3B}h_\phi\rho_0\,,\\
    \partial_k \lambda_{BBI}=&\partial_k \lambda_{BBI}\Big|_\phi+4f_{3B}h_\phi\,,\\
    \partial_k h_{B\phi}=&\partial_k h_{B\phi}\Big|_\phi+f_{2B}\lambda_{B\phi}+f_{3B}u_\phi+g_{3B}h_\phi\,,
\end{align}
where $\partial_k f\big|_\phi$ corresponds to the flow of these couplings when the dimer fields are kept fixed. By imposing that the flow of these couplings remain at zero ($\partial_k f=0$), we obtain
\begin{align}
    f_{2B}=&-\frac{\partial\lambda_{BI}\big|_\phi}{2h_\phi}+\frac{\rho_0}{4h_\phi}\partial_k\lambda_{BBI}\Big|_\phi\,,\label{sec:flow;sub:eqs;eq:f2B}\\
    f_{3B}=&-\frac{\partial_k\lambda_{BBI}\big|_\phi}{4h_\phi}\,,\label{sec:flow;sub:eqs;eq:f3B}\\
    g_{3B}=&-\frac{\partial_kh_{B\phi}\big|_\phi}{h_\phi}+\frac{\lambda_{B\phi}}{2h_\phi^2}\partial_k\lambda_{BI}\Big|_\phi\nonumber\\
    &-\frac{\lambda_{B\phi}\rho_0-u_\phi}{4h_\phi^2}\partial_k\lambda_{BBI}\Big|_\phi\,.\label{sec:flow;sub:eqs;eq:g3B}
\end{align}
The flow of the rest of the couplings is then dictated by
\begin{align}
    \partial_k u_i=&\partial_k u_i\Big|_{\phi,\rho_0}+2f_{2B}h_\phi\rho_0\,,\\
    \partial_k u_\phi=&\partial_k u_\phi\Big|_{\phi,\rho_0}+\lambda_{B\phi}\partial_k\rho_0\,,\\
    \partial_k \lambda_{B\phi}=& \partial_k \lambda_{B\phi}\Big|_{\phi,\rho_0}+2g_{3B}u_\phi\,,\\
    \partial_k h_\phi=&\partial_k h_\phi\Big|_{\phi,\rho_0}+f_{2B}u_\phi\,,
\end{align}
where we have evaluated at $\rho_B=\rho_0$. Note that although we eliminate the flow of some couplings, their effect is taken into account by the functions in Eqs.~(\ref{sec:flow;sub:eqs;eq:f2B}-\ref{sec:flow;sub:eqs;eq:g3B}). We stress that now $\lambda_{B\phi}$ carries the entire three-body physics.

Finally, the flow of the renormalization factors is simply given by
\begin{align}
    \partial_k Z_a=&2m_a\frac{\partial}{\partial\vP^2}\dot{\Gamma}^{(2)}_{a^\dag a}\Big|_{p=0,\rho_0}\,,\\
    \partial_k S_a=&i\frac{\partial}{\partial\omega_p}\dot{\Gamma}^{(2)}_{a^\dag a}\Big|_{p=0,\rho_0}\,,
\end{align}
where $a=I,\phi$, and we evaluate at zero external momentum after taking the momentum derivatives. 

\subsection{Expressions for the two-point function}
\label{app:AP;sub:Gamma2}

As explained in the previous subsection, to follow the flow of the couplings we need the two-point functions $\Gamma^{(2)}_{I^\dag I}$, $\Gamma^{(2)}_{\phi^\dag \phi}$, and $\Gamma^{(2)}_{I^\dag \phi}$. The diagrams contributing to their flow are shown in Fig.~\ref{app:AP;sub:Gamma2;fig:Gamma2}. Note that thanks to the elimination of some couplings, only a few diagrams contribute.

The explicit expressions are
\begin{widetext}
\begin{align}
    \partial_k \Gamma^{(2)}_{k,I^\dag I}=&\tilde{\partial}_k \int_q\frac{h_\phi^2}{2}\bigg[\frac{D_{B,+}(q)D_{I,-}(q+p)}{\det_B(q)\det_{I\phi,-}(q+p)}+(+\leftrightarrow -)\bigg]\,,\\
    \partial_k \Gamma^{(2)}_{k,\phi^\dag \phi}=&\tilde{\partial}_k \int_q\bigg[\frac{h^2_\phi}{2}\bigg(\frac{D_{B,-}(q)D_{\phi,-}(q+p)}{\det_B(q)\det_{I\phi,-}(q+p)}+(+\leftrightarrow -)\bigg)+\frac{2\rho_B\lambda^2_{B\phi}C_2(q)}{\det_B(q)}\bigg(\frac{D_{I,-}(q+p)}{\det_{I\phi,-}(q+p)}+(+\leftrightarrow -)\bigg)\nonumber\\
    &-\frac{\lambda_{B\phi}}{2}\frac{C_1(q)+C_2(q)}{\det_B(q)}\bigg]\,,\\
    \partial_k \Gamma^{(2)}_{k,\phi^\dag I}=&\tilde{\partial}_k\int_\Q\bigg[h_\phi\rho_B^{1/2}\lambda_{B\phi}\bigg(\frac{ D_{2,+}(q)D_{I,-}(q+p)}{\det_B(q)\det_{I\phi,-}(q+p)}+(+\leftrightarrow -)\bigg)+\frac{h_\phi^3\rho_B^{1/2}}{2}\frac{\Delta E_B(\Q)}{\det_B(q)}\bigg(\frac{1}{\det_{I\phi,-}(q+p)}-(+\leftrightarrow -)\bigg)\bigg]\,,
\end{align}
\end{widetext}
where $\tilde{\partial}_k$ is a $k$-derivative that acts on only the regulators, $(+\leftrightarrow -)$ denotes changing the signs in the subscripts of the previous expression,
\begin{align}
    C_{i}(q)&=E_{k,i}(\Q)+V_B\omega^2\,,\qquad i=1,2\\
    D_{2,\pm}(q)&=E_{k,2}(\Q)+2V_B\omega^2\pm2iS_B\omega\,,\\
    D_{B,\pm}(q)&= D_{2,\pm}(q)+E_{k,1}(\Q)\,,\\
    D_{a,\pm}(q)&=E_{k,a}(\Q)+iS_a\omega\,,\qquad a=I,\phi\\
    \text{det}_{I\phi,\pm}(q)&=D_{I,\pm}(q)D_{\phi,\pm}(q)-\rho_B h^2_\phi\,,\\
    \Delta E_B(\Q)&=E_{1,k}(\Q)-E_{2,k}(\Q)\,
\end{align}
and $\det_B$ is defined in Eq.~(\ref{app:RP;sub:eqs;eq:detB}). The regulated energies $E_{k,1}$ and $E_{k,2}$ are defined in Eqs.~(\ref{sec:RP;sub:G;eq:E1},\ref{sec:RP;sub:G;eq:E2}), and $E_{k,I}$ and $E_{k,\phi}$ in Eqs.~(\ref{sec:AP;sub:G;eq:EI},\ref{sec:AP;sub:G;eq:Ephi}). We again stress that the regulated energies depend on $\rho_B$, which is evaluated at $\rho_B=\rho_0$ only after taking the derivatives.

\subsection{RG flow examples}
\label{app:AP;sub:flows}

\begin{figure}[t!]
\centering
\includegraphics[scale=0.70]{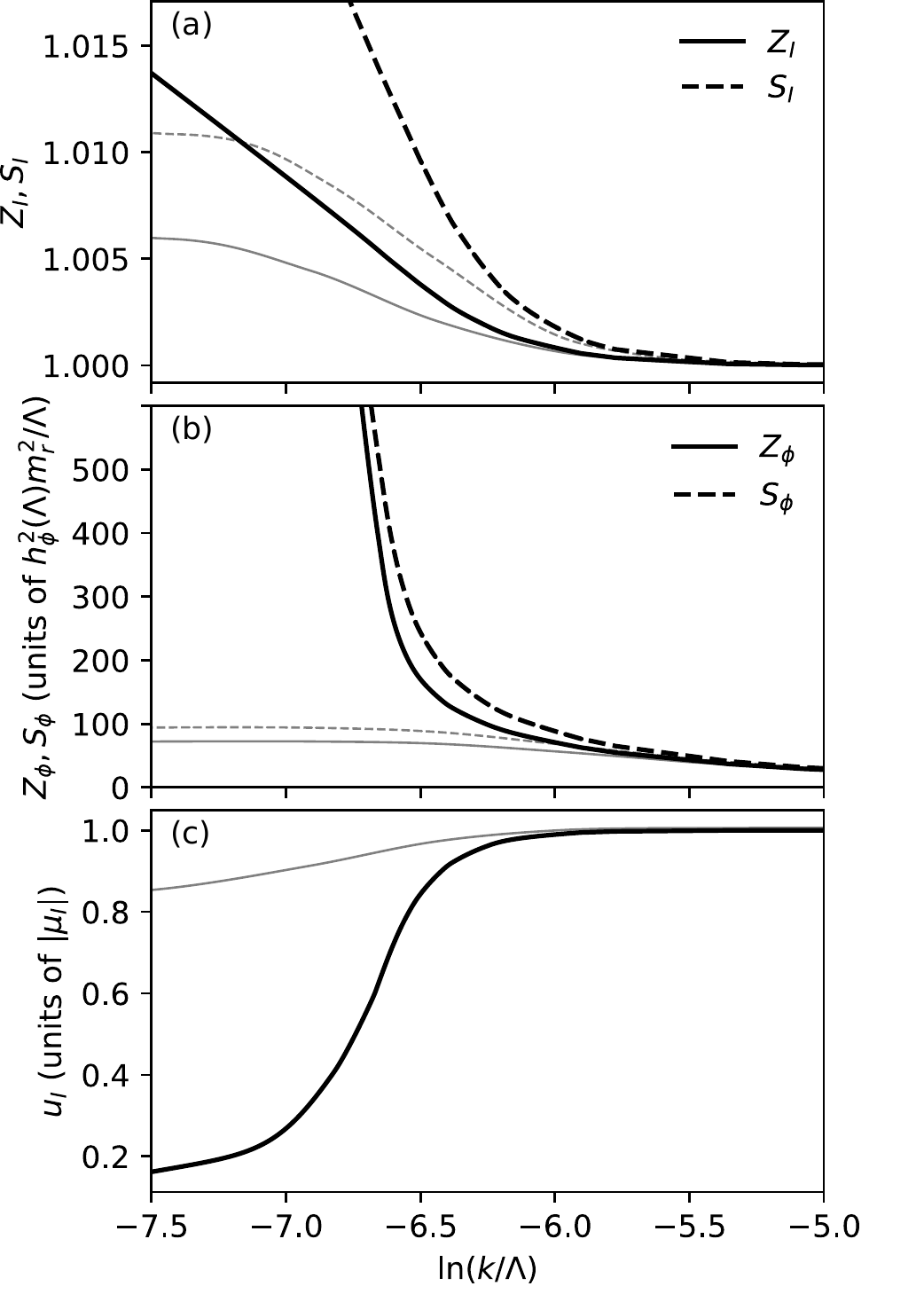}
\caption{Renormalization group flows in three dimensions at unitarity $a_{BI}^{-1}=0$ for $n^{1/3}a_{BB}=3.5\times10^{-3}$. The black lines correspond to flows at the ground state energy $\mu^*_I<0$ such that $q_{0,\pm}^*\to 0$ (\ref{sec:AP;sub:G;eq:q0pm}) for $k\to 0$. The thin gray lines correspond to an energy $\mu_I<\mu_I^*$.}
\label{app:AP;sub:flows;fig:flows}
\end{figure}

Here we show some examples of RG flows to illustrate the behavior of the couplings as functions of $k$. Fig.~\ref{app:AP;sub:flows;fig:flows} shows flows of the couplings $Z_I$, $S_I$, $Z_\phi$, $S_\phi$ and $u_I$ in three dimensions and unitarity ($a_{BI}\to\infty$) for a chosen gas parameter for the bosonic medium. The black lines are flows at the physical polaron energy $\mu_I^*$ which gives $q_{0,\pm}^*(\bm{0})\to 0$ for $k\to 0$ (see Sec.~\ref{sec:AP;sub:G}), whereas the thin gray lines correspond to flows obtained with an energy $\mu_I<\mu_I^*$. 

The renormalization factors (panels (a) and (b)) diverge at $\mu_I^*$. At this energy $\mu_I^*$ the rest of the couplings in Eq.~(\ref{sec:AP;sub:G;eq:q0pm}) vanish or saturate to finite values. For example, $u_I$ vanishes as $k$ goes to zero (panel (c)). All this results in the vanishing of $q_{0,\pm}^*(\bm{0})$ at $\mu_I^*$. On the other hand, for $\mu_I<\mu_I^*$ the renormalization factors saturate to finite values, and thus, $q_{0,\pm}^*(\bm{0})\neq 0$ for $k\to 0$. This is true for any $\mu_I\neq\mu_I^*$.

\begin{figure}[t!]
\centering
\includegraphics[scale=0.70]{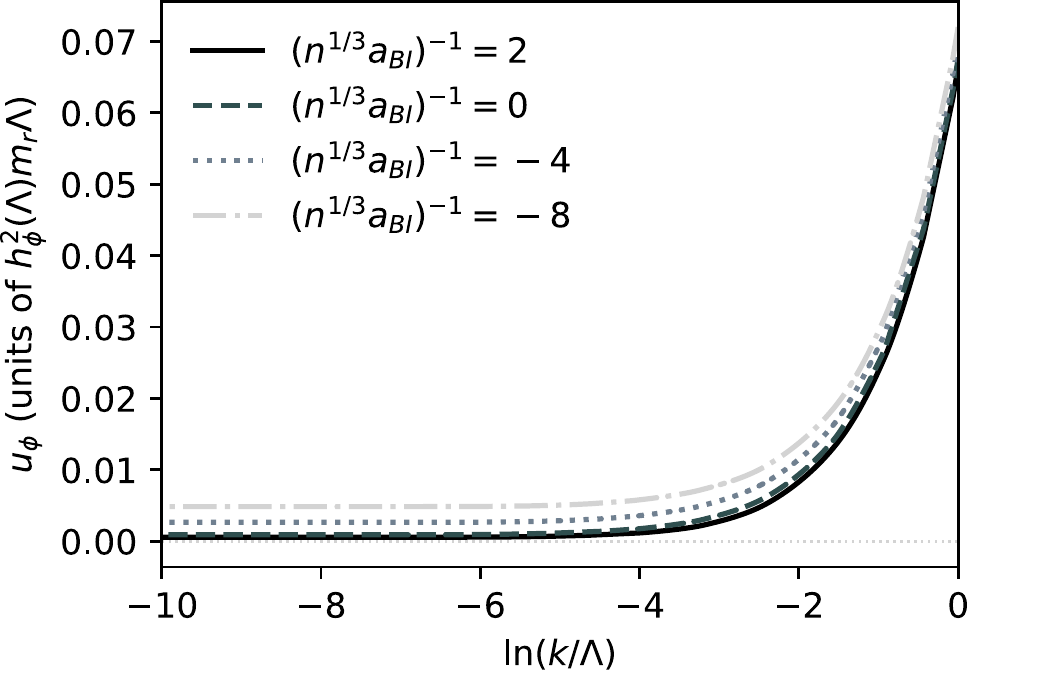}
\caption{Renormalization group flow of $u_\phi$ in three dimensions for different boson-impurity scattering lengths $a_{BI}$ for $n^{1/3}a_{BB}=3.5\times10^{-3}$. All the lines correspond to flow at the corresponding ground state energy $\mu_I^*$.}
\label{app:AP;sub:flows;fig:uphi}
\end{figure}

Fig.~\ref{app:AP;sub:flows;fig:uphi} shows the flow of $u_\phi$ in three dimensions for different values of $(n^{1/3}a_{BI})^{-1}$ at the corresponding ground state energies $\mu_I^*$. This coupling saturates to finite values. However, we observe that $u_\phi$ saturates to values closer to zero as $(n^{1/3}a_{BI})^{-1}$ increases. Our interpretation is that this signals the polaron-to-molecule crossover. As we increase the boson-impurity interaction, the molecule state dominates, and so the dimer self-energy in vacuum $u_\phi$ decreases.  Analogous flows are obtained in two dimensions.

We stress that the coupling $u_I$ can not be identified as the impurity self energy in vacuum because it contains the effect of the higher order couplings $\lambda_{BI}$ and $\lambda_{BII}$ (which have been eliminated from the flow). Therefore, we do not observe a change in $u_I$ for different $(n^{1/3}a_{BI})^{-1}$.

\section{Estimation of the effective mass in the repulsive branch}
\label{app:meff}

Because in the DE we follow the flow at zero momentum, we can not compute the complete Green's function $G(q)$, and therefore we can not extract quantities such as the effective mass and the residue~\cite{rath_field-theoretical_2013}. However, here we propose a way to estimate the effective mass in the repulsive branch, which can give us an idea of how the FRG could perform with a more robust calculation.

From the pole (\ref{sec:RP;sub:G;eq:q0}), we can naively impose at $k=0$ that
\begin{equation}
    \frac{\Q^2}{2m_I^*}=\frac{1}{S_I}\left(Z_I\frac{\Q^2}{2m_I}+u_I\right)\,,
\end{equation}
where $m_I^*$ is the estimated effective mass, in analogy to a rigorous definition~\cite{rath_field-theoretical_2013}. At the physical ground state energy $\mu_I^*$, we have that $u_I=0$. This gives the following expression
\begin{equation}
\frac{m_I^*}{m_I}=\frac{S_I}{Z_I}\bigg|_{k=0}\,,
\end{equation}
which enables us to extract the effective mass in the repulsive branch. An analogous condition was proposed in Ref.~\cite{rancon_quantum_2012} to extract effective masses with the FRG in a Bose-Hubbard model.

\begin{figure}[ht]
\centering
\includegraphics[scale=0.70]{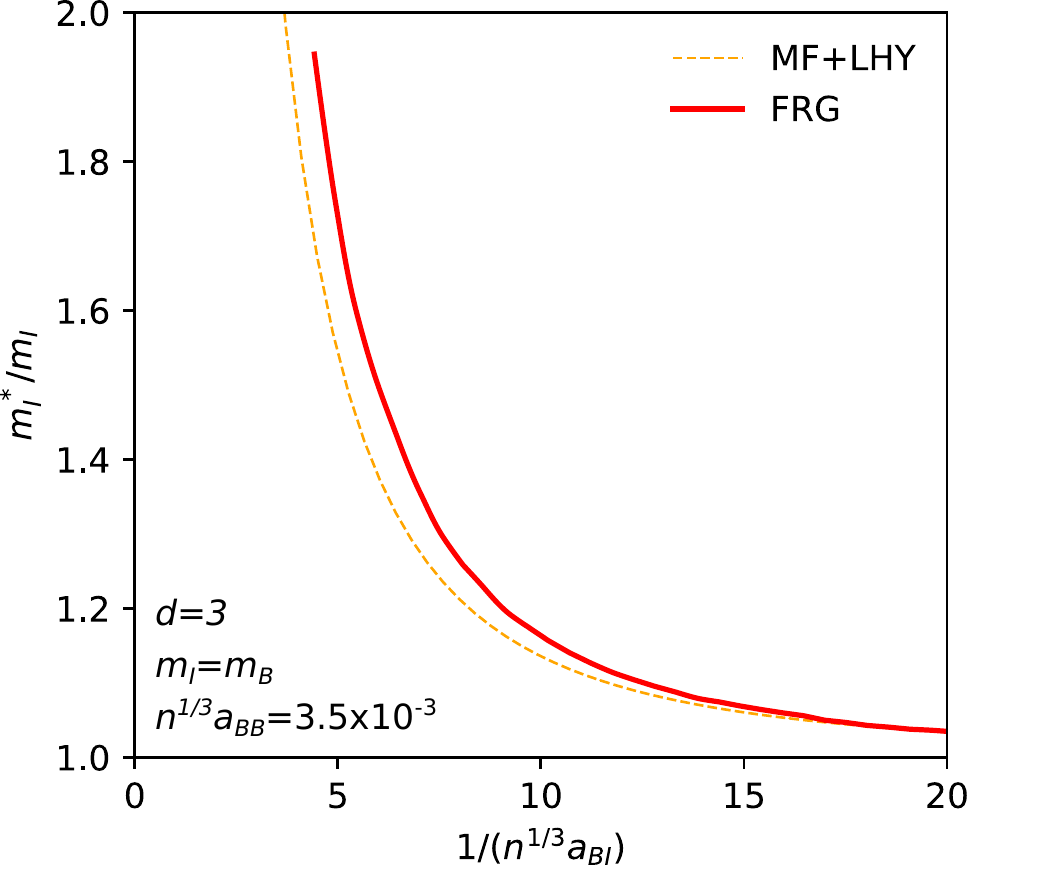}
\caption{Effective mass $m^*_I/m_I$ of the repulsive branch in three dimensions as a function of $(n^{1/3}a_{BI})^{-1}$ for $m_B=m_I$ and $n^{1/3}a_{BB}=3.5\times 10^{-3}$. The solid red line corresponds to FRG calculations, whereas the dashed orange line to the perturbative solution~(\ref{app:meff;eq:3D}). }
\label{app:meff;fig:3D}
\end{figure}

Fig.~\ref{app:meff;fig:3D} shows effective masses in three dimensions for equal boson and impurity masses. We compare with the perturbative solution for $m_B=m_I$~\cite{cucchietti_strong-coupling_2006}
\begin{equation}
    \frac{m_I^*}{m_I}=1+\frac{64}{45\sqrt{\pi}}\sqrt{na_{BB}^3}\frac{a_{BI}^2}{a_{BB}^2}\,,
    \label{app:meff;eq:3D}
\end{equation}
which was shown to give a good description compared to more sophisticated approaches~\cite{pena_ardila_impurity_2015}. Our estimate is in good agreement with the perturbative solution, showing the correct trend.

\begin{figure}[t!]
\centering
\includegraphics[scale=0.70]{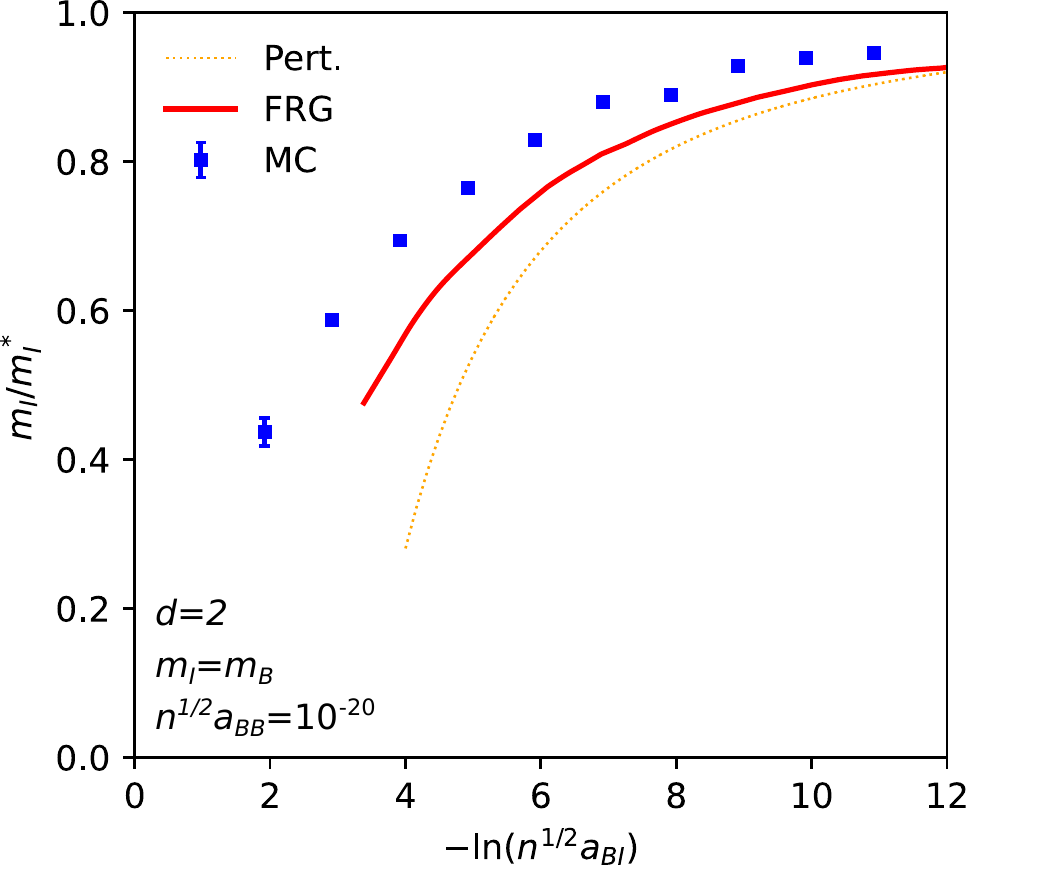}
\caption{Effective mass $m^*_I/m_I$ of the repulsive branch in two dimensions as a function of $-\ln(n^{1/2}a_{BI})$ for $m_B=m_I$ and $n^{1/2}a_{BB}=10^{-20}$. The solid red line corresponds to FRG calculations, the blue squares to MC simulations~\cite{pena_ardila_strong_2020}, and the dashed orange line to the perturbative solution~(\ref{app:meff;eq:2D}).}
\label{app:meff;fig:2D}
\end{figure}

Fig.~\ref{app:meff;fig:2D} shows effective masses in two dimensions for equal boson and impurity masses. We compare with MC simulations from Ref.~\cite{pena_ardila_strong_2020}, and with the perturbative LHY-type solution for $m_B=m_I$~\cite{pena_ardila_strong_2020}
\begin{equation}
    \frac{m_I^*}{m_I}=1+\frac{1}{2}\frac{\ln(n^{1/2}a_{BB})}{\ln^2(n^{1/2}a_{BI})}\,.
    \label{app:meff;eq:2D}
\end{equation}
We obtain reasonable agreement again with the perturbative solution and the MC results, especially with the latter. This suggests that, even within our approximation, the FRG is able to give a good description of the effective mass.

We are not able to provide a similar expression in the attractive branch where the poles have a much more complicated structure. Furthermore, because the impurity and dimer degrees of freedom are hybridized, the calculation of the effective mass is even less straightforward. We expect that in future work, we will be able to provide an accurate description by including the momentum dependence of the couplings.

\bibliography{biblio}

\end{document}